\pdfoutput=1

\documentclass[twocolumn,english]{revtex4}
\usepackage{amssymb,amsmath,graphicx}
\usepackage{color}
\usepackage[T1]{fontenc}
\usepackage{times}
\usepackage{babel}
\begin{document}


\title{
  Mesoscopic valley filter in graphene Corbino disk containing a~p-n junction 
}

\author{Dominik Suszalski}
\affiliation{Marian Smoluchowski Institute of Physics,
  Jagiellonian University, \L{}ojasiewicza 11, PL--30348 Krakow, Poland}

\author{Grzegorz Rut}
\affiliation{Marian Smoluchowski Institute of Physics,
  Jagiellonian University, \L{}ojasiewicza 11, PL--30348 Krakow, Poland}

\author{Adam Rycerz}
\affiliation{Marian Smoluchowski Institute of Physics,
  Jagiellonian University, \L{}ojasiewicza 11, PL--30348 Krakow, Poland}

\date{September 20, 2019}

\begin{abstract}
The Corbino geometry allows one to investigate the propagation of electric
current along a~p-n interface in ballistic graphene in the absence of
edge states appearing for the familiar Hall-bar geometry. 
Using the transfer matrix in the angular-momentum space we find that for
sufficiently strong magnetic fields the current propagates only in one
direction, determined by the magnetic field direction and the interface
orientation, and the two valleys, K and K', are equally occupied. 
Spatially-anisotropic effective mass may suppress one of the valley currents,
selected by the external electric field, 
transforming the system into a~mesoscopic version of the valley filter.
The filtering mechanism can be fully understood within the
effective Dirac theory, without referring to atomic-scale effects
which are significant in proposals operating on localized edge states. 
\end{abstract}

\maketitle

\section{Introduction}
One-dimensional conduction channels associated with edge states are often
considered as background for solid-state quantum information processing
not only in systems showing the quantum Hall effect
\cite{Ban18,Mro18,Bee03,Wil07,Aba07,Car11,Zim17,Can18}, 
but also in graphene \cite{Fuj96,Nak96}
or transition metal dichalcogenide nanoribbons \cite{Col18}. 
The aforementioned  nanostructures are formed of two-dimensional materials
that host an additional electronic valley degree of freedom, allowing
dynamic control and the developement of valleytronic devices \cite{Sch16},
such as the valley filter \cite{Ryc07,Ryc08}. 

The operation of early proposed valley filters in graphene, employing
the constriction with zigzag edges \cite{Ryc07} or the line defect
\cite{Gun11},
was strongly affected by atomic-scale defects \cite{Akh08} and local magnetic
order \cite{Wim08a}. 
To overcome these difficulties, alternative proposals utilizing strain-induced
pseudomagnetic fields \cite{Zha11,Jia13,Set16,Mil16,Zha17,Zha18},
disorder and curvature effects in carbon nanotubes \cite{Pal11}, 
or various types of domain walls in graphene, bilayer graphene \cite{Sch15},
or topological systems \cite{Pan15}, were put forward. 
Despite such theoretical and computational efforts the experimental
breakthrough is still missing, although some recent progress can be
noticed \cite{Gor14,Mak14,Shi15}. 
Therefore, conceptually novel mechanisms of valley
filtering are very desired. 

\begin{figure}[!t]
\centerline{
  \includegraphics[width=0.90\linewidth]{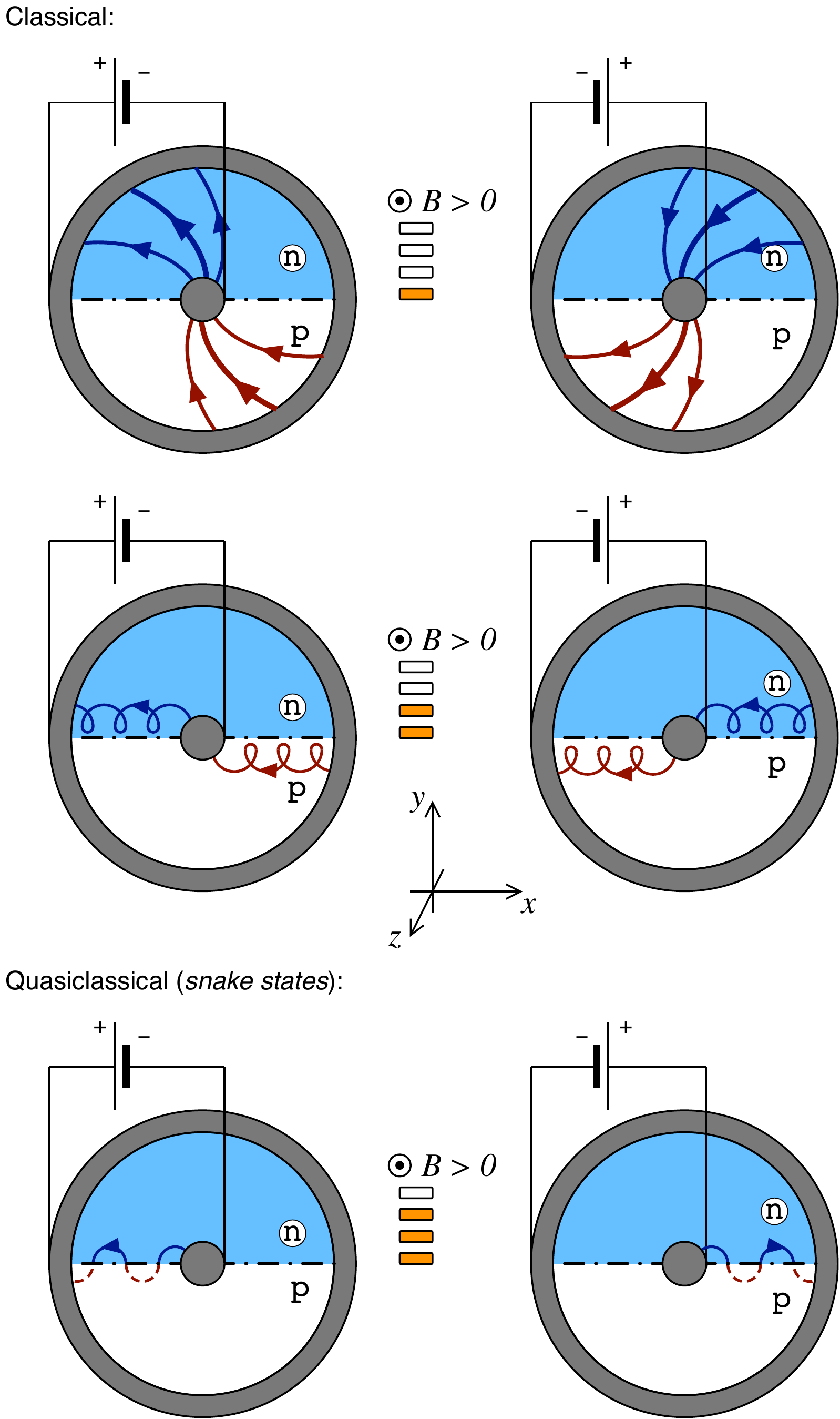}
}
\caption{ \label{snakefig}
  Classical and quasiclassical trajectories (schematic) for
  electrons (blue lines with arrows) and holes (red lines with arrows)
  in graphene Corbino disk containing
  a~p-n junction (black dash-dot line) placed in a weak (top panel),
  moderate (middle panel), 
  and strong (bottom panel) magnetic field ${\mbox{\boldmath$B$}}=(0,0,B)$,
  with $B>0$. 
  Left and right subplots correspond to the opposite polarity of a~voltage
  source driving a~current between circular leads (shadow areas). 
  The coordinate system used in the calculations is also shown. 
}
\end{figure}

In this paper, we explore the possibility of valley filtering for peculiar 
edge states mixing Landau levels from both sides of the p-n interface
in the quantum Hall regime \cite{Wil07,Aba07,Car11}.
Such unconventional edge states can be regarded as degenerate versions
of snake states, recently observed in ultraclean graphene devices
\cite{Ric15,Mak18} (see Fig.\ \ref{snakefig}). 
As the charge density is centered far from physical edges of the system,
and transport is essentially of a~{\em mesoscopic}, rather than nanoscopic,
nature (i.e., the wavefunction varies on a~length scale given by
the magnetic length $l_B=\sqrt{\hbar/eB}\gg{}a$, with $a=0.246\,$nm being
the lattice parameter; see Ref.\ \cite{Liu15}),
some of the above-mentioned obstacles in sustaining the valley polarization
of current may be overcome. 
Additionally, the Corbino geometry \cite{Ryc10,Kat10,Pet14,Kum18}
allows one to elliminate conventional edge states, making it
possible to fully control the spatial distribution of electric current via
external electric and magnetic fields. 

\begin{figure}[!t]
\centerline{
  \includegraphics[width=0.85\linewidth]{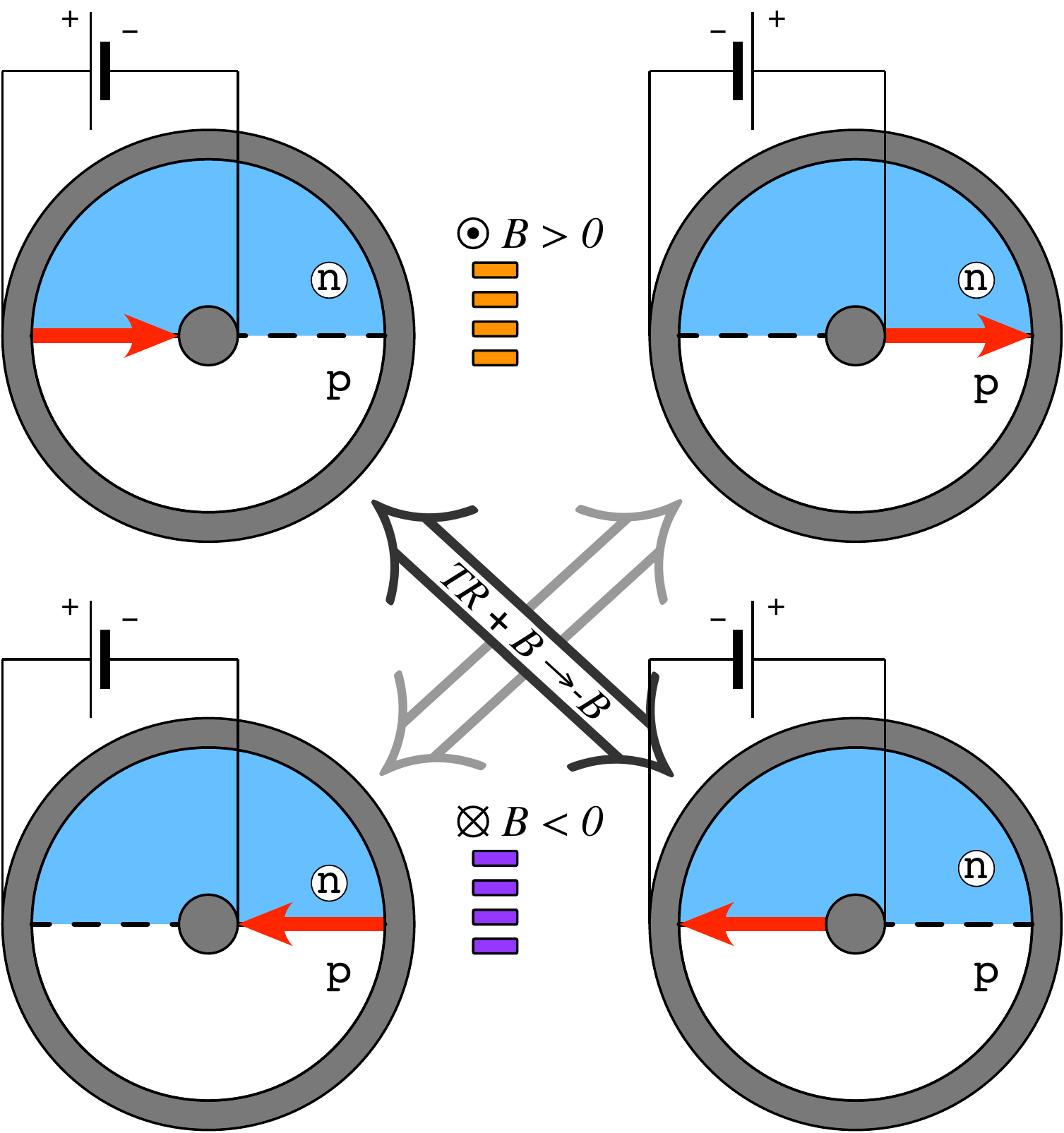}
}
\caption{  \label{qhallfig}
  Quantum Hall states propagating along a~p-n junction in the strong-field
  limit, for $B>0$ (top) and $B<0$ (bottom). 
  Diagonal double arrows indicate the system symmetry upon a~simultaneous
  time reversal and magnetic field inversion. 
}
\end{figure}

Possible classical carrier trajectories for weak-to-moderate magnetic fields
are depicted schematically in top and middle panels of Fig.~\ref{snakefig}. 
Snake states (bottom panel) cannot be understood fully clasically, as they
involve relativistic Klein tunneling through the region of an opposite
polarity. In the quantum Hall regime the current flows along one
section of the p-n interface only (see Fig.\ \ref{qhallfig}).
The physical meaning of a~``weak'', ``moderate'', or ``strong'', field 
is determined by mutual relatios between the characteristic sample length
$L\equiv{}R_o-R_i$ (with the outer disk radius $R_o$ and the inner
radius $R_i$), the magnetic length $l_B$ ($\ \propto{}B^{-1/2}$),
and the cyclotron radius $r_c$ ($\ \propto{}B^{-1}$) \cite{cycradfoo}.
In turn, the larger the disk size the lower field is required to elliminate
currents distant from the p-n interface, providing the sake of scalability
missing in previously proposed nanoscopic valley filters \cite{claquafoo}. 

\begin{figure}[!t]
\centerline{
  \includegraphics[width=0.85\linewidth]{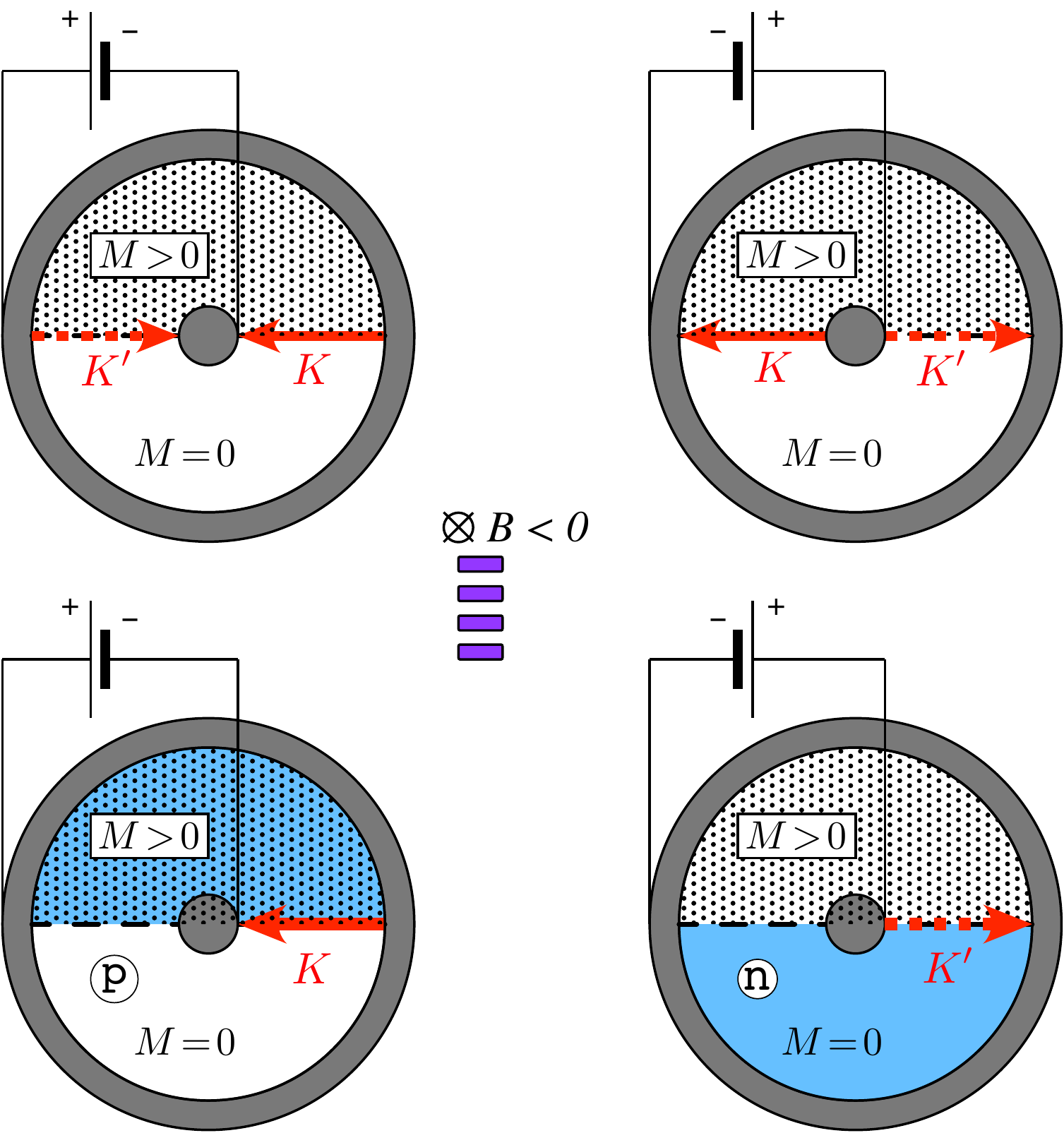}
}
\caption{  \label{pnmassfig}
  The separation of valley currents (top) and the valley polarization (bottom).
  A~constant staggered potential induces the effective mass $M>0$ in the upper
  half of the disk (dotted area). Gate electrodes (not shown) 
  tune the doping in the lower half which is undoped (top), $p$-doped
  (bottom left), or $n$-doped (bottom right). 
}
\end{figure}

We show, using the numerical transfer-matrix technique, that the presence of
a~non-uniform staggered potential, introducing the position-dependent mass
term in the effective Dirac equation for low-energy excitations \cite{Kat12},
leads to a~spatial separation of valley currents  and that the valley
polarization may be controlled by changing the gate potentials
(see Fig.\ \ref{pnmassfig}). 
Althought to set a~staggered potential one needs to initially modify
the sample on a~microscopic level, e.g., by chemical functionalization
\cite{Bou09,Hab10,Hon13} or the adsorption of hexagonal boron nitride (h-BN) 
\cite{Sac11,Yan14}, 
the operation of such a~{\em mesoscopic valley filter} is then
fully-electrostatically controlled.
We further find, that the constant magnetic field of $1\,$T 
is sufficient to obtain a~nearly perfect polarization in the disk of
a~$400\,$nm diameter.
What is more, the filter operation can be directly attributed to a~peculiar
combination of symmetry breakings for the Dirac Hamiltonian:
The mass term breaks the effective time-reversal symmetry in a~single valley
({\em symplectic symmetry}), whereas the magnetic field breaks the
{\em true} time-reversal symmetry (involving the valley exchange).
Together, these two symmetry-breaking factors lead to the inequivalence
of valleys, providing an opportunity to produce nonequilibrium valley
polarization of current. 

The paper is organised as follows:
In Sec.\ II, we briefly present the effective Dirac theory and the
transfer matrix approach to the scattering problem in the angular-momentum
space (adjusted to the Corbino-disk symmetry). 
In Sec.\ III, we discuss our numerical results concernig the current
distribution and valley filtering in the presence of external electromagnetic
field and the staggered potential. The conclusions are given in Sec.\ IV.

\section{Model and methods}

\subsection{The effective Dirac equation}
Let us start by considering a ring-shaped sample, characterized by the inner 
radius $R_{i}$ and the outer radius $R_{o}$, surrounded by metallic contacts 
modelled by heavily-doped graphene areas (we set $R_{o}=4R_{i}=200\,$nm for 
all systems considered in the paper). 
Since we focus on {\em smooth} (or long-range) disorder, the intervalley
scattering
can be neglected and one can consider the single-valley Dirac equation 
\begin{multline}
  \left(\xi\pi_{x}\sigma_{x}+\pi_{y}\sigma_{y}\right)\psi\left(r,\phi\right) =
  \\
  \left[E-\mathcal{V}\left(r,\phi\right) - \mathcal{M}\left(r,\phi\right)
  \sigma_{z}\right]\psi\left(r,\phi\right),\ \label{eq:dirac} 
\end{multline}
where $\xi=1$ ($-1$) is the valley index for $K$ ($K'$) valley,
$\sigma_\alpha$ (with $\alpha=x,y,z$) is the Pauli matrix,
$\pi_{\alpha}/v_{F}=\left(-i\hbar\partial_{\alpha}+eA_{\alpha}\right)$ 
is the gauge-invariant momentum operator with $v_{F}\approx10^{6}\,$m/s
the Fermi velocity, $E$ denotes the Fermi energy, and
$\mathcal{V}\left(r,\phi\right)$ and $\mathcal{M}\left(r,\phi\right)$
are position-dependent electrostatic potential energy and mass (respectively)
in polar coordinates $\left(r,\phi\right)$. 
We choose the symmetric gauge $\boldsymbol{A}=
\frac{B}{2}\left(-y,x\right)$ with a uniform magnetic field $B$.
Furthermore, $B\neq0$ for the disk area ($R_{i}<r<R_{o}$) only; inside
the leads ($r<R_{i}$ or $r>R_{o}$) we simply set $B=0$, as the value of
$B$ becomes irrelevant in the high-doping limit (see e.g.\ Ref.\ \cite{Ryc10}). 

In the case of a system with cylindrical symmetry (namely, $\mathcal{V}$
and $\mathcal{M}$ being $\phi$-independent), the Hamiltonian in 
Eq.\ (\ref{eq:dirac}) commutes with the angular-momentum operator, 
$L_{z}=-i\hbar\partial_{\phi}+\xi\hbar\sigma_{z}/2$, and the wavefunction
can be expressed as a product of radial and angular parts
\begin{equation}
\psi_{l}\left(r,\phi\right)=\varphi_{l}\left(\phi\right)\theta_{l}\left(r\right)\equiv e^{i(l-\xi\sigma_{z}/2)\phi}\left[\begin{array}{c}
\theta_{A,l}\left(r\right)\\
\theta_{B,l}\left(r\right)
\end{array}\right],\label{eq:pseudospindecomp}
\end{equation}
where $l$ is an half-odd integer, and $A$ ($B$) labels 
the upper (lower) spinor element.

\subsection{Mode-matching in the angular-momentum space
  \label{subsec:Method-of-approach}}

To solve the scatering problem numerically we simplify here, for the case
of a~monolayer, the method earlier developed for the Corbino disk in bilayer
graphene \cite{Rut16}. 

If $\mathcal{V}$ or $\mathcal{M}$ in Eq.\ (\ref{eq:dirac}) is $\phi$-dependent 
the cylindrical symmetry is broken, however, one still can employ
the angular-momentum eigenfunctions to represent a~general solution 
as a~superposition
\begin{equation}
\psi\left(r,\phi\right)=\sum_{k}\psi_{k}\left(r,\phi\right), 
\end{equation}
with $\psi_{k}\left(r,\phi\right)$ given by Eq.\ (\ref{eq:pseudospindecomp})
[see also {\it Appendix~A}]. 

Substituting the above into Eq.\ (\ref{eq:dirac}) we obtain 
\begin{multline}
\sum\limits_{k} \hbar v_{F}\left[\partial_{r}-f^{k}\left(r\right)\right]\psi_{k}\left(r,\phi\right)  = \\
\sum\limits_{k} \xi\left\{ i\sigma_{x}\left[E-\mathcal{V}\left(r,\phi\right)\right]-\sigma_{y}\mathcal{M}\left(r,\phi\right)\right\} \psi_{k}\left(r,\phi\right),
\label{eq:diracpolar} 
\end{multline}
where $f^{k}\left(r\right)=\left[\xi\left(k/r+l_{B}^{-2}r/2\right)\sigma_{z}-1/\left(2r\right)\mathbb{I}_{2\times2}\right]$, the magnetic length 
$l_{B}=\sqrt{\hbar/\left(eB\right)}$, and $\mathbb{I}_{n\times n}$
is the $n\times n$ identity matrix. Multiplication over the conjugate
angular wavefunction $\varphi_{l}^{*}\left(\phi\right)$ and subsequent
integration over the polar angle $\phi$ leads to
\begin{multline}
\left[\partial_{r}-f^{l}\left(r\right)-i\sigma_{x}\frac{\xi E}{\hbar v_{F}}\right]\theta_{l}\left(r\right) = \\
\sum_{k}\xi\left[-i\sigma_{x}\mathbb{V}_{lk}\left(r\right)-\sigma_{y}\mathbb{M}_{lk}\left(r\right)\right]\theta_{k}\left(r\right), \label{eq:dirK}
\end{multline}
with 
\begin{equation}
\mathbb{V}_{lk}\left(r\right)=\frac{1}{2\pi}\intop_{0}^{2\pi}d\phi\left[\mathcal{V}\left(r,\phi\right)/\left(\hbar v_{F}\right)\right]e^{i(k-l)\phi},\label{eq:matPot}
\end{equation}
and
\begin{equation}
\mathbb{M}_{lk}\left(r\right)=\frac{1}{2\pi}\intop_{0}^{2\pi}d\phi\left[\mathcal{M}\left(r,\phi\right)/\left(\hbar v_{F}\right)\right]e^{i(k-l)\phi}.\label{eq:matPotM}
\end{equation}
(Notice that the angular dependence of $\mathcal{V}$ or $\mathcal{M}$
introduces the mode-mixing in our scattering problem.) 

The general solution of Eq. (\ref{eq:dirK}) can be written as
a vector $\boldsymbol{\theta}\left(r\right)=\left[\theta_{l_{min}}^{A}\left(r\right),...,\theta_{l_{max}}^{A}\left(r\right),\theta_{l_{min}}^{B}\left(r\right),...,\theta_{l_{max}}^{B}\left(r\right)\right]^{T}$, 
with cutoff angular-momentum quantum numbers $l_{min}$ and $l_{max}$. 
(Hereinafter, $L=l_{max}-l_{min}+1$ is the total number of transmission
modes.)
Subsequently, one can write 
\begin{multline}
\left\{ \left[\partial_{r}+1/\left(2r\right)\right]\mbox{\ensuremath{\mathbb{I}}}_{2L\times2L}-\xi\sigma_{z}\otimes\mathbb{L}\right\} \boldsymbol{\theta}\left(r\right) = \\
\xi\left[i\frac{E}{\hbar v_{F}}\sigma_{x}\otimes\mathbb{I}_{L\times L}-i\sigma_{x}\otimes\mathbb{V}-\sigma_{y}\otimes\mathbb{M}\right]\boldsymbol{\theta}\left(r\right), \label{eq:dirK-1}
\end{multline}
where $\mathbb{A}\otimes\mathbb{B}$ is the Kronecker product of
matrices $\mathbb{A}$ and $\mathbb{B}$, and the diagonal matrix
\begin{multline}
  \mathbb{L} = \mbox{diag}
  \left(\frac{l_{min}}{r} + \frac{r}{2l_{B}^{2}},
  \frac{l_{min}\!+\!1}{r} + \frac{r}{2l_{B}^{2}}, \right. \\
  \left.  \dots, \frac{l_{max}}{r} + \frac{r}{2l_{B}^{2}} \right).
\end{multline}

Once the scattering matrix is determined (see {\it Appendix~A} for details),
transport properties of the system can be calculated 
within the Landauer-B\"{u}ttiker formalism in the linear-response regime
\cite{Lan70,But92}. 
In particular, the electrical conductance and valley polarization are
given by 
\begin{equation}
  G = G_{0}\sum_{\xi=\pm{}1}\mbox{Tr}\,\mathbf{T}_\xi, \ \ \ \ \ \ 
  {\cal P} = \frac{\mbox{Tr}\,\mathbf{T}_{\xi=1}-\mbox{Tr}\,\mathbf{T}_{\xi=-1}}{%
        \mbox{Tr}\,\mathbf{T}_{\xi=1}+\mbox{Tr}\,\mathbf{T}_{\xi=-1}}, 
  \label{eq:landauer}
\end{equation}
where $G_{0}=2e^{2}/h$, the prefactor $2$ marks the spin degeneracy
(we neglect the Zeeman effect \cite{zeemafoo}), 
and $\mathbf{T}_\xi=\mathbf{t}_{\xi}^{\dagger}\mathbf{t}_{\xi}$ with
$\mathbf{t}_\xi$ being the transmission matrix for one valley. 
We further neglect the electron-electron interaction and electron-phonon
coupling,
which is a~common approach to nanosystems in monolayer graphene close to
the Dirac point, as the scattering processes associated with these many-body
effects are usually slower than the ballistic-transport processes
\cite{Mue09,Luc18}.

The matrix  $\mathbf{t}$ is also employed when calculating the radial
current density, which is given by 
\begin{equation}
\label{jraver}
  j_{r}\left(r,\phi\right) = ev_{F}\sum_{l}
  \psi_{l}^{\dagger}\!\left(r,\phi\right)
  J_r({\phi})\,\psi_{l}\left(r,\phi\right),
\end{equation}
with the radial current density operator 
\begin{equation}
\label{Jroper}
J_r({\phi})=\xi\sigma_{x}\cos\phi+\sigma_{y}\sin\phi, 
\end{equation}
and $\psi_{l}\left(r,\phi\right)=\sum_{k}\left(\mathbf{t}_\xi\right)_{l,k}\,e^{i(k-\xi\sigma_{z}/2)\phi}\left(1,\xi\right)^{T}/\sqrt{r}$
being the transmitted wavefunction in the outer contact ($r>R_{o}$). 
The matrix element $\left(\mathbf{t}_{\xi}\right)_{l,k}$ denotes the
transmission probability amplitude from channel $k$ to $l$. 
Similarly, the cartesian components of the current density
${\mbox{\boldmath{$j$}}}(r,\phi)=\left(\,j_x,j_y\,\right)$ 
are calculated by replacing the operator $J_r$ in Eq.\ (\ref{jraver}) by 
\begin{equation}
\label{Jxyoper}
  J_x = \xi\sigma_x \ \ \ \ \text{or } \ \ \ \ J_y=\sigma_y \ \ \ \ 
  \text{(respectively)}. 
\end{equation}

\section{Quantum transport in crossed electric and magnetic fields} 

\subsection{Definitions}

\begin{figure}[!t]
\includegraphics[width=8cm]{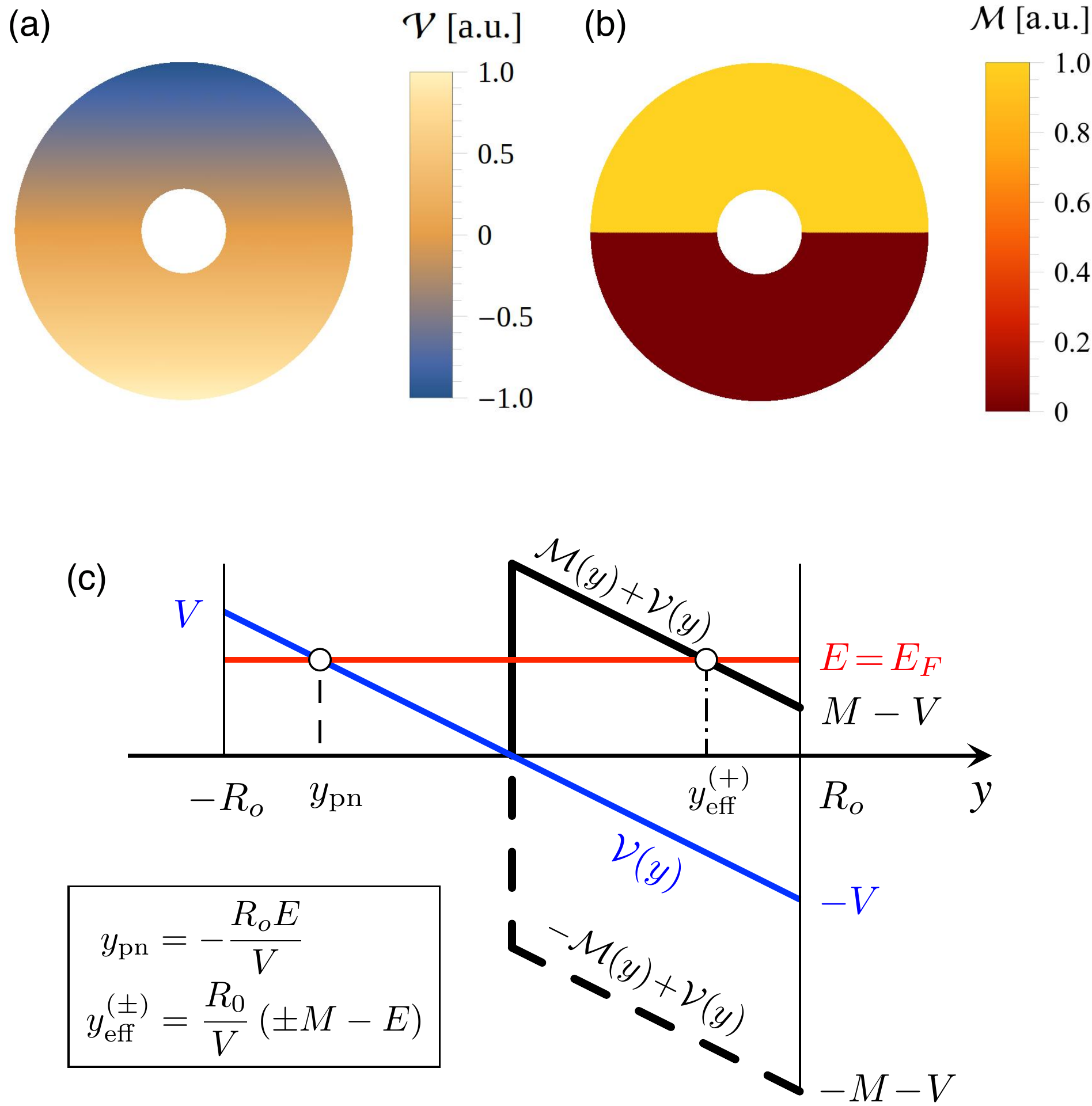}
\caption{ \label{fig:masa}
  (a) Electrostatic potential energy $\mathcal{V}\left(r,\phi\right)$
  given by Eq.\ (\ref{eq:epot}) with $\phi_V=0$, 
  corresponding to a~homogeneous electric field
  $\mathbf{E}=\left(0,\mathcal{E}\right)$.
  (b) The mass term  $\mathcal{M}\left(r,\phi\right)$ given by
  Eq.\ (\ref{eq:masa}) with $\phi_M=0$ (i.e., $\mathcal{M}=M\neq{}0$
  in the upper half of the disk only).
  (c) Crossections of the electrostatic potential energy ${\cal V}(y)$
  (blue line) and the effective potentials ${\cal V}(y)\pm{}{\cal M}(y)$
  (black solid or dashed line) along the disk diameter: $x=0$, $-R_o\leqslant
  y\leqslant{}R_o$, for some $V,M>0$.  Red line marks the Fermi energy $E$.
  The expressions for $y_{\rm pn}$ (the position of a~p-n interface), following
  from ${\cal V}(y_{\rm pn})=E$, and $y_{\rm eff}^{(\pm)}$ (for which ${\cal V}\pm
  {\cal M}=E$) are also given.
  (The inner-lead edges, $y=\pm{}R_i$, are omitted for clarity.) 
}
\end{figure}

In order to study a~role of the p-n junction in quantum transport through
graphene-based Corbino disk, we choose the electrostatic potential energy
as follows
\begin{equation}
  \label{eq:epot}
  \mathcal{V}\left(r,\phi\right)=
  -e\mathcal{E}r\sin\left(\phi-\phi_{V}\right),
\end{equation}
where $\mathcal{E}$ is the electric field (we further define $V\equiv{}
e\mathcal{E}R_{o}$) and the angle $\phi_{V}$ defines the crystallographic
orientation of  the p-n interface \cite{Luk07,Shy09}. 
Furthermore, we investigate how the transport is affected
by the mass term 
\begin{equation}
\label{eq:masa}
  \mathcal{M}\left(r,\phi\right)=
  M\Theta\left(\phi-\phi_{M}\right)\Theta\left(\pi+\phi_{M}-\phi\right), 
\end{equation}
with the angle $\phi_{M}$ specifying the mass arrangement,  
and $\Theta\left(x\right)$ being the Heaviside step function. 
The mass term given by Eq.\ (\ref{eq:masa}) is restricted to a~half of
the disk,  $\phi\in\left[\phi_{M},\pi+\phi_{M}\right]$, see
Fig.\ \ref{fig:masa}.
In the heavily-doped contact regions, 
$\mathcal{V}\left(r,\phi\right)=\mathcal{M}\left(r,\phi\right)=0$. 

It is worth to mention that we have also considered other functional forms
of the mass term, inluding $\mathcal{M}\left(r,\phi\right)$ smoothly
varying with the distance from a~p-n junction, always finding a~parameter
range in which the valley-filtering mechanism that we describe was highly
efficient. Even for a~simple model given by Eq.\ (\ref{eq:masa}), 
changing the Fermi energy ($E$) allows one to shift a~p-n
interface ($E-\mathcal{V}\left(r,\phi\right)=0$) with respect to the mass
boundary, leading to a~rich phase diagram discussed later in this section.

The specific forms of the potential energy $\mathcal{V}\left(r,\phi\right)$
and the mass term $\mathcal{M}\left(r,\phi\right)$, given by Eqs.\
(\ref{eq:epot}) and (\ref{eq:masa}), lead to the matrix elements

\begin{equation}
\mathbb{V}_{lk}=
  -\delta_{\left|l-k\right|,1}\frac{Vr\pi}{hv_{F}R_{o}}
  \,i^{k-l}\exp\left[i\left(k-l\right)\phi_{V}\right], 
\end{equation}
and 
\begin{equation}
\mathbb{M}_{lk}=\frac{Mr}{hv_{F}R_{o}}\begin{cases}
\pi & \mbox{\mbox{if  \ensuremath{\left|l-k\right|=1}}},\\
i\frac{1-(-1)^{k-l}}{k-l}\mbox{e}^{i\left(k-l\right)\phi_{M}}
 & \mbox{if  \ensuremath{\left|l-k\right|\neq1}}.
\end{cases}
\end{equation}

\subsection{Quantum Hall regime in the massless case}

\begin{figure}[!t]
\includegraphics[width=8cm]{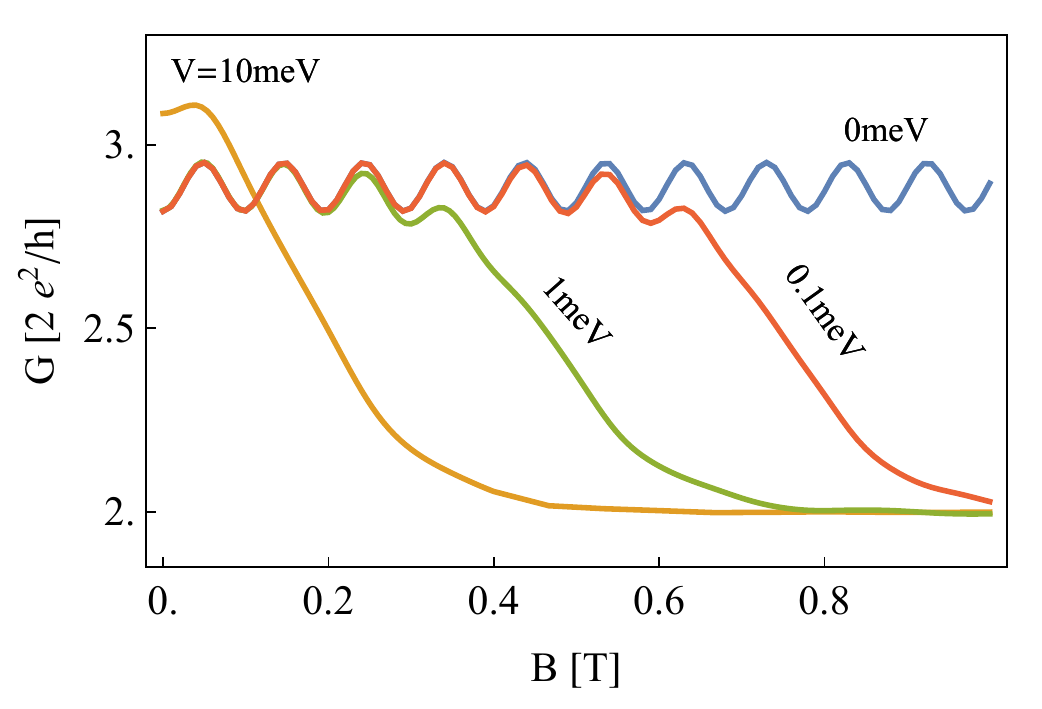}

\caption{\label{fig:plotGB}
Magnetoconductance of the Corbino disk with $R_{o}=4R_{i}=200\,$nm, $E=M=0$,
and different values of the in-plane electric field ${\cal E}$
(quantified by $V \equiv{} e{\cal E}R_o$). 
Notice that the conductance ($G$) approaches the one quantum value 
($4e^{2}/h$) for any $V\neq{}0$ at sufficiently high magnetic field ($B$). }

\end{figure}

We consider now the case of $M=0$ in Eq.\ (\ref{eq:masa}).
The Fermi energy is set as $E=0$ and thus the p-n interface overlaps
with the disk diameter ($y_{\rm pn}=0$) for any $V\neq{}0$.

For moderate values of the electric field ($|V|<10\,$meV) and weak magnetic
fields the magnetoconductance behavior 
is same as in a~case without the p-n junction \cite{Ryc10}, 
see Fig.\ \ref{fig:plotGB}.
The increase of $G$ at weak magnetic fields, visible for $|V|=10\,$meV, 
indicates the system is close to the ballistic transport regime. 
This occurs when the (position-dependent) cyclotron diameter
$2{}r_{c}(0,y)\gtrsim{}R_{o}\!-\!R_{i}$, enhancing vertical currents along the
classical trajectories (cf.\ the top panel in Fig.\ \ref{snakefig}). 
For our choice of the parameters, the cyclotron radius, 
\begin{equation}
r_{c}(x,y)=\frac{|E-{\cal V}(x,y)|}{eBv_F}, 
\end{equation}
is bounded by 
$r_{c}(0,y)\geqslant{}(R_{i}/R_{o})|V|/\left(eBv_{F}\right)$ along the vertical
diameter ($x=0$) and for $R_{i}\leqslant{}|y|\leqslant{}R_{o}$.

%

\begin{figure*}[!t]
\includegraphics[width=0.8\linewidth]{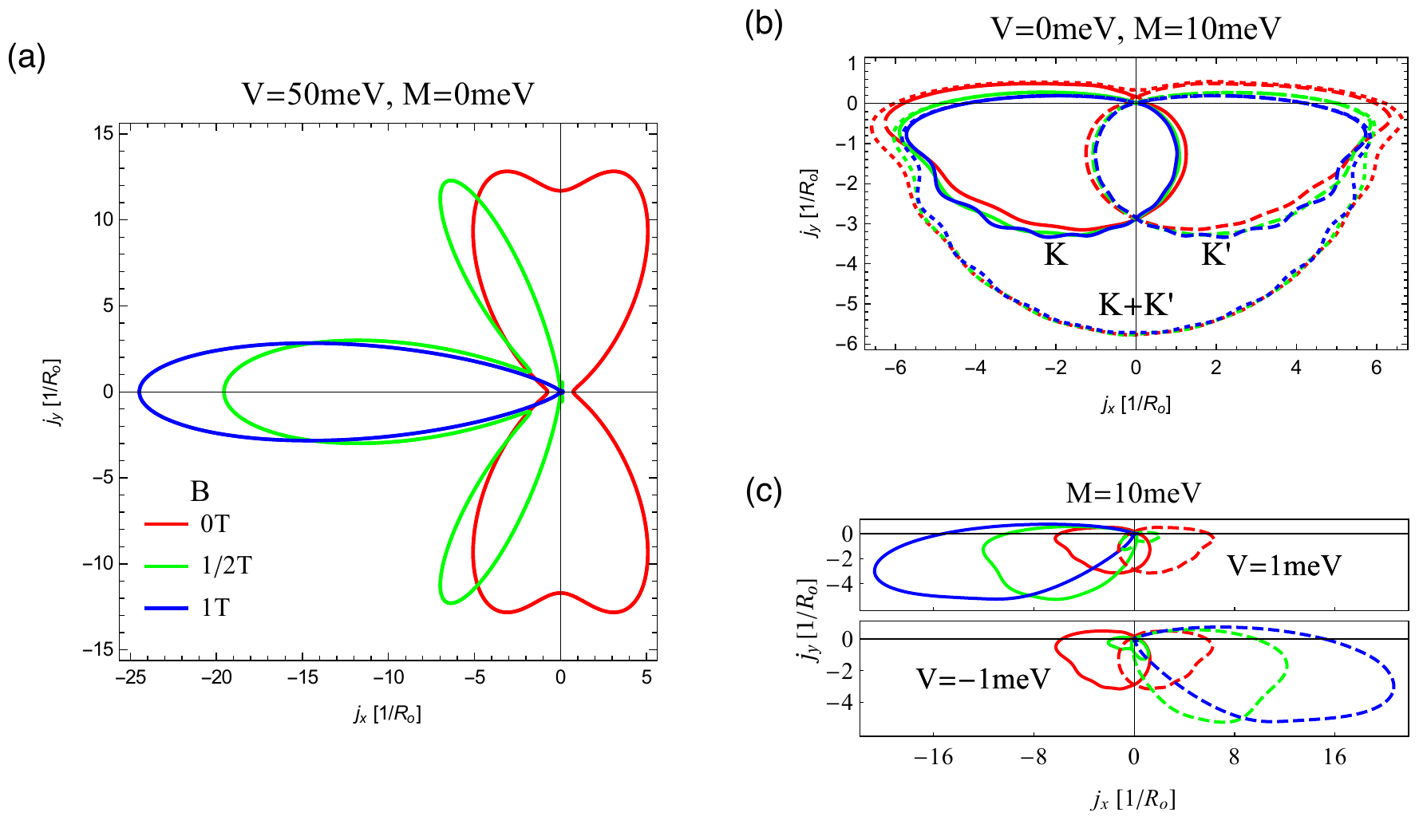}
\caption{ \label{fig:RhoJ3pan}
  Parametric plots of the current density $\left.\mbox{\boldmath$j$}
  (r,\phi)\right|_{r=R_{o}}\equiv{}(j_x,j_y)$ [see Eq.\ (\ref{Jxyoper})], 
  where $0\leqslant{}\phi<2\pi$,
  for the same setup as in Fig.\ \ref{fig:plotGB} with the electrostatic
  potential and mass magnitudes ($V$ and $M$) varied between the subplots.
  Line colours (same for each subplot) mark different magnetic fields:
  $B=0$ (red), $B=1/2\,$T (green), and $B=1\,$T (blue). 
  (a) For $M=0$, the results are identical for both valleys. 
  A~relatively large value of $V=50\,$meV elliminates the pseudodiffusive
  charge-transport regime. 
  At $1\,$T field, the distribution resembles the analytical result for an
  infinite plane with the p-n junction, see Fig.\ \ref{fig:Current-density}
  in {\it Appendix~B}. 
  (b) For $V=0$ and $M=10\,$meV, the separation of valley currents appears. 
  Solid and dashed lines correspond to distinct contributions from the two
  valleys ($K$, $K'$); dotted lines depict the current density summarized
  over the valleys ($K+K'$).
  (c) For $M=10\,$meV and $V=\pm{}1$meV (top/bottom panel), 
  polarity of the p-n junction allows one to select one of the valley currents
  and suppress the other. 
  (The summarized current density is omitted for clarity.)
}
\end{figure*}

Another apparent feature of the data presented in Fig.\ \ref{fig:plotGB}
is a~rapid conductance drop, occuring for any $V\neq{}0$ at sufficiently
high field. Unlike in a~uniformely-doped disk out of the charge-neutrality
point, where $G$ vanishes in the~high-field limit \cite{Ryc10}, here 
$G$ approaches the value of $4e^{2}/h$ (i.e., the {\em conductance quantum}
with spin and valley degeneracies) 
signalling the crossover from pseudodiffusive to quantum-hall transport
regime. The limiting value of $G$ reproduces the experimental result of
Ref.\ \cite{Pet14}, and can be easily explained by analysing symmetries of
the Dirac theory \cite{Bee08}.

A~bit more detailed view of the effect is provided with the evolution of
angle-dependent current density at the outer disk edge ($r=R_{o}$) with
increasing field, presented in Fig.\ \ref{fig:RhoJ3pan}(a). 
We choose a~high electric field ($V=50\,$meV) to ensure the system undergoes
a~crossover directly from ballistic to quantum-Hall transport regime,
as the contribution from evenescent waves is negligible.
For $B=0$ (red line) the current flows in directions along which the doping
is extremal, namely, $\phi=\pm{}\pi/2$.
For higher fields the transport is dominated by a~single direction,
for which $\mathcal{V}\left(r,\phi\right)=0$ (i.e.,  $\phi=\pi$),
with some secondary currents at $|\phi|\lesssim\pi/2$ visible for
$B=1/2\,$T (green line), and vanising for $B=1\,$T (blue line). 
This picture is in agreement with the results of previous theoretical studies
(see Ref.\ \cite{Bee08} and {\it Appendix~B} for details).

As the magnetic length at $1$~Tesla field 
$l_B(B\!=\!1\,\text{T})\approx{}26\,$nm is still comparable with the system
size (in particular, the inner radii $R_{i}=50\,$nm), the transport cannot be
understood classically or quasiclasically. 
Therefore, several features depicted
schematically in Fig.\ \ref{snakefig} (such as the orbits in the middle panel)
have no correspondants in numerical results presented
in Fig.\ \ref{fig:RhoJ3pan}(a).
However, an apparent asymmetry of the current distribution for $B\neq{}0$
is directly linked to the left-right mirror symmetry breaking, also present
in the classical level: Both the trajectories and quantum-hall edge states are
symmetric upon a~simulaneous left-right reflection and the field inversion
(cf.\ Fig.\ \ref{qhallfig}); the same applies to the voltage-source polarity
(or time) reversal combined with the magnetic field inversion.

\subsection{Mass term and the valley filter operation}

So far, we have put $M=0$ in Eq.\ (\ref{eq:masa}) and the transport
characteristics were identical for both valleys (K and K').
A different picture emerges in the system with nonzero and
spatially-varying mass term (the $M\neq{}0$ case).
Our simplified model, in which the mass is present only in the upper half
of the system (see Fig.\ \ref{fig:masa}), already allows to demonstrate
the mesoscopic valley-filtering mechanism. 
In this subsection, we present the central results of the paper, 
providing a~quantitative description of the effects depicted
schematically in Fig.~\ref{pnmassfig}.

Quite surprisingly, even at zero electric and magnetic fields the currents
corresponding to different valleys are well separated
(see Fig.\ \ref{fig:RhoJ3pan}(b)).
This can be interpreted as a~zero-doping version the edge-state formation
(the Fermi energy is fixed at $E=0$). 
As the mass opens a~band gap in the upper half of the disk ($0<\phi<\pi$),
there are no extended states available, and the current is pushed away towards
the lower half ($-\pi<\phi<0$). 
In turn, the border between areas with ${\cal M}=0$ and ${\cal M}\neq{}0$
plays a~role of an artificial edge of the system (notice that the p-n junction
is absent for $V=0$).
The {\em total} current distribution (dotted lines in Fig.\
\ref{fig:RhoJ3pan}(b))
is approximately uniform in the lower half of the disk (as this
part is in the pseudodiffusive charge-transport regime), with some local
maxima for $\phi\approx{}0$ and $\phi\approx{}-\pi$, signaling contributions
from the zero-energy edge states.
The emergence of such states is well-described in graphene literature,
see e.g.\ Ref.\ \cite{Wim09}; their analogs in bilayer graphene in 
a~position-dependent perpendicular electric field were also discussed
\cite{Sch15}.
A~basic reasoning why electrons in different valleys
prefer opposite directions of propagation is given in {\it Appendix~C}. 

A~direct link between the valley polarization of current and the direction
of propagation for zero-energy edge states leads to the spatial separation
of valley currents, which is apparent even in our relatively small system,
for which the role of evenescent waves is still significant (and manifests
itself by a~nonozero current density for any $-\pi<\phi<0$).

%

%

\begin{figure}[!t]
\includegraphics[width=8.5cm]{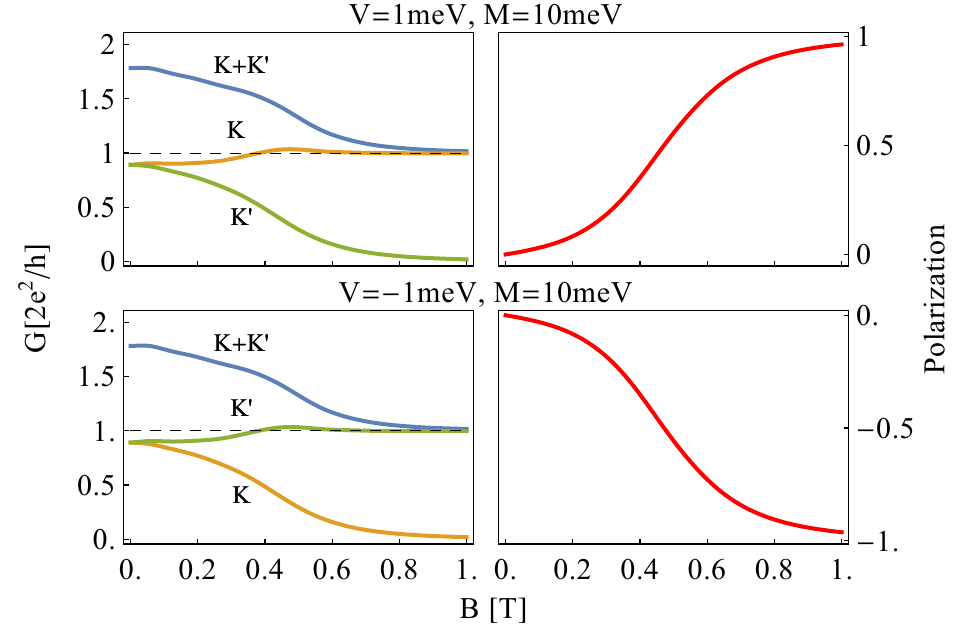}

\caption{\label{fig:ConMass}
  Conductance (left) and valley polarization (right) defined by
  Eq.\ (\ref{eq:landauer}) for the Corbino disk with both the mass term
  and the p-n interface displayed as functions of the magnetic field.
  Remaining system parameters are same as in Fig.\ \ref{fig:RhoJ3pan}(c).
}
\end{figure}

Next, the valley-filtering mechanism is demonstrated by creating the
p-n interface in a~presence of the mass term ($V\neq{}0$, $M\neq{}0$).
Fig.\ \ref{fig:RhoJ3pan}(c) shows a~strong suppression of one of the
valley currents in relatively weak electric and magnetic fields
(and the valley is selected by a~{\em sign} of $V$), provided that
the mass term is sufficently strong.
The valey polarization ${\cal P}$ gradually increases with the magnetic
field, becoming almost perfect for $B=1\,$T (see Fig.\ \ref{fig:ConMass}).

\begin{figure}[!t]
\includegraphics[width=8cm]{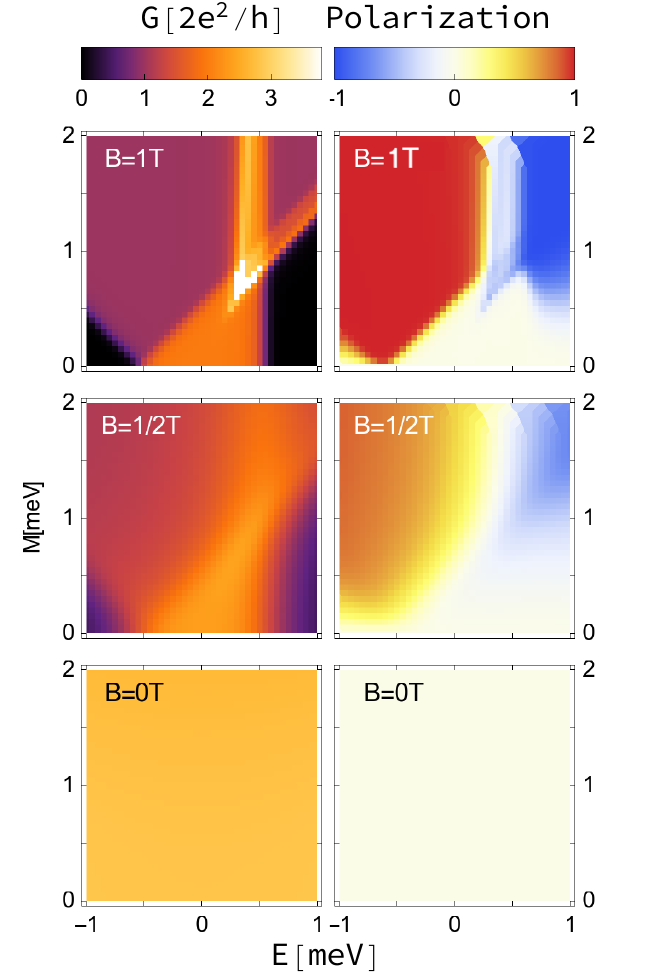}
\caption{\label{fig:przegladPlaska}
Conductance (left) and valley polarization (right) for $V=1\,$meV
as functions of the Fermi energy ($E$) and the mass term ($M$).
The value of magnetic field ($B$) is varied between the panels. 
Remaining system parameters are same as in Fig.\ \ref{fig:plotGB}. 
}
\end{figure}

\begin{figure}[!t]
\includegraphics[width=0.8\linewidth]{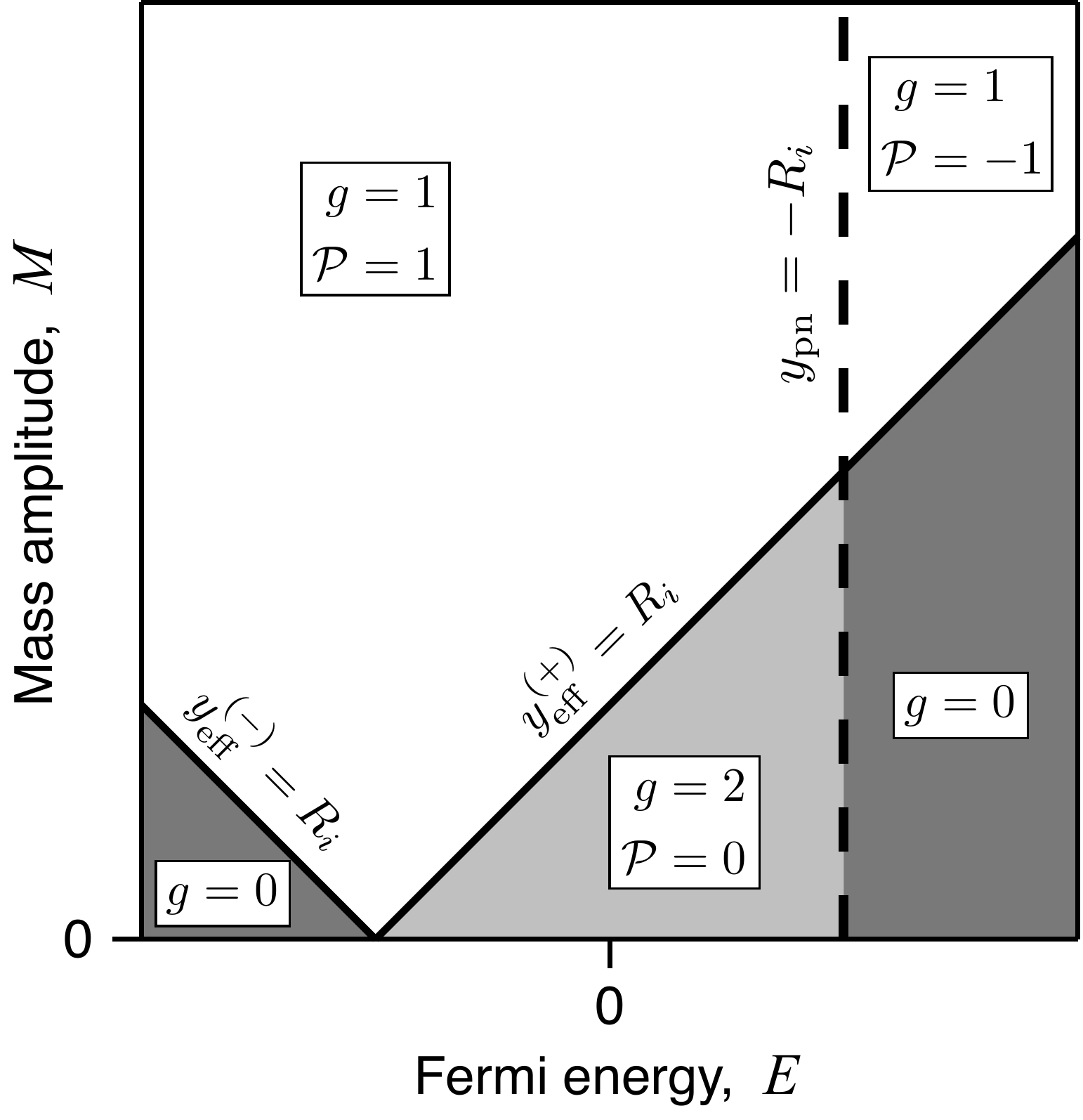}
\caption{ \label{fig:PhDiag1T}
  Sketch of the high-field ``phase diagram'' for some $V>0$
  (see top panel in Fig.\ \ref{fig:przegladPlaska}) with distinct
  regions characterized by dimensionless conductance $g=G/G_0$ and
  the polarization ${\cal P}$, 
  with the boundaries given by mutual relation between the Fermi energy 
  and the potentials ($\,{\cal V}\!\pm{}\!{\cal M}$ and ${\cal V}\,$) at
  the inner disk edge: $y_{\rm eff}^{(\pm)}=R_i$ (solid lines) and $y_{\rm pn}=-R_i$
  (dashed line). [See also Fig.\ \ref{fig:masa}(c).] 
}
\end{figure}

The operation of our valley filter is characterized in details by the
numerical results presented in Fig.\ \ref{fig:przegladPlaska}, where we
have fixed $V=1\,$meV, and visualized the transport characteristics in 
the Fermi energy-mass ($E$--$M$) parameter plane, 
for three selected values of the magnetic field ($B=0$, $1/2$, and $1\,$T).
Notice that varying $E$ corresponds to a~vertical shift of the p-n interface;
in particular, for $E=VR_{i}/R_{o}=0.25\,$meV we have $y_{\rm pn}=-R_{i}$
(cf.\ Fig.\ \ref{fig:masa}) and the p-n interface is a~tangent line to the inner
disk edge at the lower (i.e., mass-free) half. 
At zero magnetic field, the density maps shown in bottom panels are perfectly
uniform, and no valley polarization is visible. 
For higher fields, distinct regions of the "phase diagram" are formed,
including the unpolarized highly-conducting region
($G\approx{}2G_0$, ${\cal P}\approx{}0$) at the central-bottom part of each
subplot, the two polarized highly-conducting regions
($G\approx{}G_0$, ${\cal P}\approx{}\pm{}1$) near the upper corners,
and the two tunneling regions ($G\approx{}0$, ${\cal P}\approx{}0$)
near the lower corners. 
At $1\,$T field (top panels), the boundaries between above-mentioned regions
are already well-developed.

Some further insights into relations connecting the diagram structure and 
characteristic features of the effective potential profile, 
${\cal V}(r,\phi)\pm{}{\cal M}(r,\phi)\equiv{}{\cal V}(y)\pm{}{\cal M}(y)$
in Eq.\ (\ref{eq:dirac}), are given in Fig.\ \ref{fig:PhDiag1T}.
In brief, the boundaries between regions on the $E$--$M$ diagram can be
attributed to the situations when the p-n line is a~tangent to the inner
disk edge at the mass-free part, $y_{\rm pn}=-R_i$ (vertical dashed line),
or when the Fermi energy is equal to the effective potential along a~tangent
line to the inner disk edge at the nonzero mass part,  $y_{\rm eff}^{(\pm)}=R_i$
(diagonal solid lines).
The schetch of Fig.\ \ref{fig:PhDiag1T} corresponds to the high-field limit,
in which $l_B\ll{}R_i$ and varying $E$ may lead to an abrupt
switching between the regions. In a~finite-field situation
(see Fig.\ \ref{fig:przegladPlaska}), finite widths of quantum Hall states
result in blurs (and shifts) of the boundaries, with a~general trend to
expand the unpolarized highly-conducting region with decreasing $B$. 

Numerous experimental realizations of a~non-uniform mass in monolayer
graphene \cite{Bou09,Hab10,Hon13,Sac11,Yan14} suggest to focus on
a~constant and relatively large $M\gg{}1\,$meV.
In such a~case, the magnetic field of $B=1\,$T allows one to control the valley
polarization of current independently by tuning the Fermi energy ($E$)
or by reversing the p-n junction polarity ($V\rightarrow{}-V$). 

It is also worth stressing, that high valley polarization remains unaffected
when the p-n interface is moved by a~distance of $\Delta{}y\approx{}R_{i}=
50\,$nm away from the mass boundary, allowing us to coin the term of 
mesoscopic valley filter.

\section{Conclusions}
We have demonstrated, as a~proof of principle, that the Corbino disk
in monolayer graphene modified such that the mass term in effective Dirac
equation is present in a~half of the disk (leading to the energy gap of
$\gtrsim{}1\,$meV) may act as a~highly efficient valley filter, when
placed in crossed electric and magnetic fields inducing a~p-n interface
close to the mass-region boundary.
Although introducing the mass term involves a~microscopic modification
of a~sample, the output (valley) polarization of current may be controlled
electrostatically in constant magnetic field, 
alternatively by: (i) inverting the p-n junction polarity, 
or (ii) shifting the p-n line with respect to the mass boundary by tuning
a~global doping of a~sample. 
The magnetic field of $1\,$T is sufficient to obtain the polarization better
than $99$\% for the device size (namely: the outer disk diameter) of
$400\,$nm. 

An additional interesting feature of the system is that the currents
belonging to different valleys are spatially separated, flowing in opposite
directions along the p-n interface. In the absence of a~p-n interface,
there are two equal currents propagating along the mass boundary; 
in-plane electric field amplifies one of these currents and supresses
the other. The filtering mechanism is directly linked to global symmetry
breakings of the Dirac Hamiltonian, and therefore we expect it to be robust
against typical perturbations in real experiments. 

For instance, 
the operation of mesoscopic valley filter which we have desribed should
not be noticeably affected by the long-range (or smooth) impurities, as they
generally do not introduce the intervaley scattering \cite{Bar07,Ryc12}.
(In contrast, short-range impurities mix the valleys and may restore the
equilibrium valley occupation.) Recent experimental works on
ultraclean graphene p-n junctions \cite{Ric15,Mak18} allow us to believe that
such systems, accordingly modified to induce a~position-dependent 
quasiparticle mass, may also act as highly-efficient
mesoscopic valley filters.

\section*{Acknowledgements}

Discussions with Piotr Witkowski are appreciated. 
The work was supported by the National Science Centre of Poland (NCN)
via Grant No. 2014/14/E/ST3/00256. Computations were partly performed
using the PL-Grid infrastructure.

\appendix

\section{Transfer matrix approach}

A general wavefunction corresponding to the $l$-th transmission channel 
is given by a linear combination of two linearly-independent spinor functions 
\begin{equation}
  \theta^{l}\left(r\right)=
  a_{1}^{l}\theta_{1}^{l}\left(r\right)+a_{2}^{l}\theta_{2}^{l}\left(r\right),
\end{equation}
where $a_{\alpha}^{l}$ ($\alpha=1,2$) are arbitrary complex amplitudes and
$\theta_{\alpha}^{l}\left(r\right)=\left[
  \theta_{\alpha,A}^{l},\theta_{\alpha,B}^{l}\right]^{T}\left(r\right)$
is a normalized spinor function with $A$ and $B$ being the sublattice indices.
The normalization has to be carried out in such a~way that the total current
remains constant (i.e., $l,\alpha$--independent). 
To satisfy this condition, we write down the current density for the
$l$-th transmission channel 
\begin{equation}
  \vec{j}_{l}=ev_{F}\left[\theta^{l}\left(r\right)\right]^{\dagger}\cdot
  \left[
    \xi\sigma_{x}\cos\left(\varphi\right)+\sigma_{y}\sin\left(\varphi\right)
  \right]
  \cdot\theta^{l}\left(r\right).
\end{equation}
In principle, it is sufficient to normalize only the wavefunctions
in the leads since the relation between them (nammely: between the incoming,
the transmitted, and the reflected wavefunction) ultimately defines matrices
$\mathbf{r}$ and $\mathbf{t}$. Current conservation guarantees that amplitudes
$r_{mn}$ and $t_{mn}$ preserve the probabilistic interpretation.
Therefore, a~direct normalization for the wavefunctions in the sample area
is not essential for the successful mode matching.

Next, 
it is convenient to present a~complete set of wavefunctions as a vector
with each element corresponding to a different transmission channel.
Since only a limited number of channels contributes significantly to the
quantum transport, one can look for a~truncated solution by 
introducing the cutoff-transmission channels $l_{min}$ and $l_{max}$ such that
$l\in\left[l_{min},l_{max}\right]$.
The total number of transmission channels, $M=\left(l_{max}-l_{min}+1\right)$,
is chosen to be large enough to reach the convergence. 
In such a~notation, we can write
\begin{equation}
  \boldsymbol{\theta}\left(r\right)=
  \mathbb{M}\left(r\right)\left(\begin{array}{c}
  \boldsymbol{a}_{1}\\
  \boldsymbol{a}_{2}
  \end{array}\right),\label{eq:phimat}
\end{equation}
where $\mathbb{M}\left(r\right)$ is a $2M\times{}2M$ matrix,
$\boldsymbol{a}_{\alpha}= 
\left[a_{\alpha}^{l_{min}},\ldots,a_{\alpha}^{l_{max}}\right]^{T}$.
The explicit form of matrix $\mathbb{M}\left(r\right)$ will be presented
later. The notation of Eq.\ (\ref{eq:phimat}) is convenient when 
dealing with a~system with mode mixing introduced by a~position-dependent
potential. 

We are primarily interested in a~relation between the two sets of amplitudes
defining wavefunctions at different radii, say: $r$ and $R_{i}$.
Such a~relation can be written introducing a propagator
$\mathbb{U}\left(r,R_{i}\right)$, 
\begin{eqnarray}
\boldsymbol{\theta}\left(r\right) & = &
\mathbb{U}\left(r,R_{i}\right)\boldsymbol{\theta}\left(R_{i}\right). 
\label{eq:prop}
\end{eqnarray}
The propagator $\mathbb{U}\left(r,R_{i}\right)$ can be found by substituting
Eq.\ (\ref{eq:prop}) into Eq.\ (\ref{eq:dirK-1}) from the main text
(the Dirac equation). 
The resulting equation takes the following form 
\begin{equation}
\partial_{r}\mathbb{\mathbb{U}}\left(r,R_{i}\right)=
\mathbb{A}\left(r\right)\mathbb{\mathbb{U}}\left(r,R_{i}\right),
\label{eq:rroz}
\end{equation}
with an initial condition
$\mathbb{\mathbb{U}}\left(R_{i},R_{i}\right)=\mathbb{I}_{2M\times2M}$.
The matrix $\mathbb{A}\left(r\right)$ in Eq.\ (\ref{eq:rroz})
carries the complete information about the potential and the mass term
in the system.

Formally, Eq.\ (\ref{eq:rroz}) defines $2M$ independent systems of $2M$
ordinary differential equations, each of which describing a~column
in the matrix $\mathbb{U}\left(r,R_{i}\right)$. 
We have employed a~fixed-step
explicit Runge Kutta method of the $4$-th order \cite{Bur11}.
Both the step-size as well as the number of transmission channels $M$ are
adjusted to reach the numerical convergence; in practise, these parameters
depend on the system size, as well as on the magnetic field, in
an approximately linear manner similarly as in the case of bilayer graphene
(see Ref.\ \cite{Rut16}). 

Once the propagator for the sample area $\mathbb{U}\left(R_{o},R_{i}\right)$ 
is determined,  we can translate it onto a~transfer matrix, connecting
the wavefunctions in the leads with wavefunctions in the sample area, 
via the mode-matching 
\begin{eqnarray}
\boldsymbol{\phi}^{L}\left(R_{o}\right) & = & \boldsymbol{\phi}^{S}\left(R_{o}\right)\nonumber \\
 & = & \mathbb{\mathbb{U}}\left(R_{o},R_{i}\right)\boldsymbol{\phi}^{S}\left(R_{i}\right)\nonumber \\
 & = & \mathbb{\mathbb{U}}\left(R_{o},R_{i}\right)\boldsymbol{\phi}^{L}\left(R_{i}\right),
\end{eqnarray}
where $\boldsymbol{\phi}^{L}\left(R_{o}\right)=\mathbb{M}_{L}\left(r\right)\boldsymbol{a}$. 
As the doping in the leads is set to infinity, 
the matrix $\mathbb{M}_{L}\left(r\right)$ can be presented as a Kronecker
product $\mathbb{M}_{L}\left(r\right)=\mathbb{B}\left(r\right)\otimes
\mathbb{I}_{M\times M}$
(we have omitted the phase constants as they are insignificant when
calculating the transport properties), where
\begin{equation}
\mathbb{B}\left(r\right)=\frac{1}{\sqrt{r}}\left[\begin{array}{cc}
1 & 1\\
\xi & -\xi
\end{array}\right].
\end{equation}
Columns in the matrix $\mathbb{B}\left(r\right)$ represents independent
wavefunctions, corresponding to different directions of propagation
(incoming and outgoing waves).
The transfer matrix is thus given by
\begin{equation}
\mathbb{T}=\mathbb{M}_{L}^{-1}\left(R_{o}\right)\mathbb{\mathbb{U}}\left(R_{o},R_{i}\right)\mathbb{M}_{L}\left(R_{i}\right).\label{eq:transfer}
\end{equation}

Finally, the transmission properties of the system can be obtained
by retriving the scattering-matrix elements from $\mathbb{T}$.
The transfer matrix can be expressed by blocks of the scattering matrix
as follows
\begin{equation}
\mathbb{T}=\left[\begin{array}{cc}
\left(\mathbf{t}^{\dagger}\right)^{-1} & \mathbf{r}'\cdot\left(\mathbf{t}'\right)^{-1}\\
-\left(\mathbf{t}'\right)^{-1}\cdot\mathbf{r}' & \left(\mathbf{t}'\right)^{-1}
\end{array}\right],
\end{equation}
where $\mathbf{t}$ and $\mathbf{r}$ are the transmission and reflection
matrix (respectively) for a~wavefunction incoming from the inner lead; 
similarly, 
$\mathbf{t}'$ and $\mathbf{r}'$ are the transmission and reflection matrix
for a~wavefunction incoming from the outer lead.

\section{Solutions for an inifite graphene plane
  \label{sec:Appendix-B:-Wavefunctions}}

The clear asymmetry of a~current propagating along the p-n junction
in the quantum Hall regime (see Fig.\ \ref{fig:RhoJ3pan}(a) in the main text)
illustrate an intrinsic feature that is not related to the Corbino geometry. 
In this Appendix we derive analytically the eigenfunctions for the
low-energy Hamiltonian of graphene in crossed electric and magnetic fields
\begin{equation}
\label{eq:cartH}
H=\left(\begin{array}{cc}
-e\mathcal{E}x & \xi\pi_{x}-i\pi_{y}\\
\xi\pi_{x}+i\pi_{y} & -e\mathcal{E}x
\end{array}\right),
\end{equation}
where $\pi_{\alpha}=-i\hbar\partial_{\alpha}+eA_{\alpha}$ with the
Landau gauge $\mathbf{A}=\left(0,Bx\right)$, and the mass term
is neglected for simplicity.
[Notice that the electrostatic potential energy term in Eq.\ 
(\ref{eq:cartH}) corresponds to $\phi_V=\pi/2$ in Eq.\ (\ref{eq:epot}).]
It is clear now that the Hamiltonian (\ref{eq:cartH}) is invariant under
the time reversal combined with the magnetic field inversion, namely
\begin{equation}
  H(-\xi,-B) = {\cal T}_{\xi}H(\xi,B){\cal T}_{\xi}^{-1}, 
\end{equation}
where ${\cal T}_{\xi}=\sigma_0{\cal C}$ is a~single-valley time reversal
operator with ${\cal C}$ denoting complex conjugation.
(In the four-component notation, the full time reversal is
${\cal T}=\tau_x\otimes{\cal T}_\xi$, where $\tau_x$ is the Pauli matrix
acting on valley degrees of freedom.)  

Due to the translation symmetry in the $y$-direction, 
$H$ (\ref{eq:cartH}) also commutes with $-i\hbar\partial_{y}$ and thus
we can choose the wavefunction as $\Psi\left(x,y\right)=
\Phi\left(x\right)\mbox{exp}(ik_{y}y)$,
with the wavenumber $k_{y}$, reducing the scattering problem to
a~single-dimensional one.
The corresponding Dirac equation reads 
\begin{eqnarray}
\left[\begin{array}{cc}
\frac{e\mathcal{E}}{i\hbar v_{F}}x & \xi\partial_{x}+k_{y}+\frac{eB}{\hbar}x\\
\xi\partial_{x}-k_{y}+\frac{eB}{\hbar}x & \frac{e\mathcal{E}}{i\hbar v_{F}}x
\end{array}\right]\Phi\left(x\right) & = \label{DiracEqB} \\
\frac{iE}{\hbar v_{F}}\Phi\left(x\right).\nonumber 
\end{eqnarray}
One can further simplify the above equation introducing the dimensionless
variable  $\chi=l_{B}^{-1}x+l_{B}k_{y}$, where $l_{B}=\sqrt{\hbar/e|B|}$
is the magnetic length. Without loss of generality, we can suppose that $B>0$.  
Eq.\ (\ref{DiracEqB}) can now be written as
\begin{equation}
\left[\begin{array}{cc}
-\gamma\chi & -i\left(\xi\partial_{\chi}+\chi\right)\\
-i\left(\xi\partial_{\chi}-\chi\right) & -\gamma\chi
\end{array}\right]\Phi\left(x\right)=
\varepsilon\Phi\left(x\right), \label{eq:HamX}
\end{equation}
where we have defined $\gamma=el_{B}^{2}\mathcal{E}/\left(\hbar v_{F}\right)$
and $\varepsilon=l_{B}\left[E/\left(\hbar v_{F}\right)-\gamma k_{y}\right]$. 
When considering an inifinite graphene plane we can choose (without
loosing the generality) the zero Fermi energy ($E_{F}=0$), what leads to 
\begin{equation}
\varepsilon=-l_{B}\gamma k_{y}. \label{eq:eps_i_ky}
\end{equation}

Following Refs.\ \cite{Per07,Nat14}, we find the solutions of
Eq.\ (\ref{eq:HamX}) by solving an auxiliary eigensystem
\begin{equation}
\label{eq:calH}
  \mathcal{H}\varphi\left(x\right)=
  \varepsilon^{2}\varphi\left(x\right)
\end{equation}
for the operator
\begin{equation}
  \mathcal{H}=\varepsilon\left(H+\tilde{H}\right)-H\tilde{H},
\end{equation}
where $\tilde{H}=\sigma_{z}H\sigma_{z}$, which is chosen such
that each eigenfunction of $\mathcal{H}$ satisfies Eq.\ (\ref{eq:HamX})
as well.
Eq.\ (\ref{eq:calH}) can rewritten as follows 
\begin{equation}
\left(\begin{array}{cc}
-O_{-} & i\xi\gamma\\
i\xi\gamma & -O_{+}
\end{array}\right)\left(\begin{array}{c}
u\\
v
\end{array}\right)=\varepsilon^{2}\left(\begin{array}{c}
u\\
v
\end{array}\right), \label{eq:Hjpjn}
\end{equation}
where $O=2\varepsilon\gamma\chi+\varepsilon^{2}-\left(1-\gamma^{2}\right)\chi^{2}+\partial_{\chi}^{2}\pm\xi$,
and $u,v$ are spinor elements of the wavefunction $\varphi\left(x\right)$. 

We can now write down the fourth-order differential equation for $u$,
namely
\begin{equation}
\gamma^{2}u+O_{+}O_{-}u=0, \label{eq:JnJp}
\end{equation}
being equivalent to the set of two second-order equations 
\begin{multline}
  \sqrt{1-\gamma^{2}}\,u_{\pm} = \\
  \pm\left[2\gamma\varepsilon\chi+\varepsilon^{2}-
  \left(1-\gamma^{2}\right)\chi^{2}+\partial_{\chi}^{2}\right]u_{\pm}.
\end{multline}
The solutions are
\begin{equation}
u_{\pm}=a^{\pm}\mbox{D}_{-\left(1\pm1\right)/2+w}\left(\rho\right)+b^{\pm}\mbox{D}_{-\left(1\mp1\right)/2-w}\left(i\rho\right),
\end{equation}
where $\mbox{D}_{\nu}\left(x\right)$ is the parabolic cylinder function
\cite{Abr65}, 
$\rho=\sqrt{2}\left(\chi-\gamma^{2}\chi-\gamma\varepsilon\right)\left(1-\gamma^{2}\right)^{-3/4}$,
$w=\varepsilon^{2}\left(1-\gamma^{2}\right)^{-3/2}/2$, and $a^{\pm}$, $b^{\pm}$
are arbitrary constants.
Since we are interested in square-integrable wavefunctions, we set $b^{\pm}=0$.
Using Eq.\ (\ref{eq:Hjpjn}), we obtain the full form of the spinor function
\begin{align}
\left(\begin{array}{c}
u_{\pm}\\
v_{\pm}
\end{array}\right) &= a^{\pm}\mbox{D}_{-\left(1\pm1\right)/2+w}\left(\rho\right)
\nonumber \\
&\times{}\left[\begin{array}{c}
1\\
i\left(1\pm\xi\sqrt{1-\gamma^{2}}\right)/\gamma
\end{array}\right].
\end{align}

Both the solutions $\left(u_{+},v_{+}\right)^{T}$ and
$\left(u_{-},v_{-}\right)^{T}$, as well as their arbitrary linear combination,
satisfy Eq.\ (\ref{eq:Hjpjn}). 
Therefore, we construct an eigenfunction of Eq.\ (\ref{eq:calH}),
corresponding to an eigenvalue $\varepsilon^2>0$, by taking \cite{Mac83}
\begin{equation}
\label{eq:varphinon0}
\varphi_{\varepsilon^2>0}(x) \equiv 
\left(\begin{array}{c} u \\ v \end{array}\right)=A\left[c\left(\begin{array}{c} u_{+} \\ v_{+} \end{array}\right)+\left(\begin{array}{c} u_{-} \\ v_{-} \end{array}\right)\right],
\end{equation}
where $c=\varepsilon\left[1+\xi\sqrt{1-\gamma^{2}}\right]/\left[\sqrt{2}\gamma\left(1-\gamma^{2}\right)^{3/4}\right]$, 
and $A$ is the normalization constant 
\begin{equation}
A=\sqrt{\frac{\left(1-\gamma^{2}\right)^{1/4}\gamma^{2}\left(4\sqrt{\pi}l_{B}\right)^{-1}}{\mbox{\ensuremath{\Gamma}}\left[1+w\right]\left(1+\xi\sqrt{1-\gamma^{2}}\right)}}.
\end{equation}

The case of $\varepsilon^2=0$ (the {\em zero mode}) is slightly different, and
it is instructive to consider it separately. 
The corresponding solution of Eq.\ (\ref{eq:calH}) 
reads
\begin{equation}
\varphi_{\varepsilon^2=0}\left(x\right)=
Ce^{-\sqrt{1-\gamma^{2}}\chi^{2}/2}\left(\begin{array}{c}
i\frac{\xi\sqrt{1-\gamma^{2}}-1}{\gamma}\\
1
\end{array}\right), \label{eq:HolyGrail}
\end{equation}
with 
\begin{equation}
C=\sqrt{\frac{\gamma^{2}\left(1-\gamma^{2}\right)^{1/4}}{2\sqrt{\pi}\left(1+\xi\sqrt{1-\gamma^{2}}\right)l_{B}}}.
\end{equation}

In a~general case, the normalization of $\varphi(x)$ leads also to a~discrete
spectrum of eigenvalues
\begin{equation}
\varepsilon_n^{2}=2\left(1-\gamma^{2}\right)^{3/2}\,n, 
\end{equation}
with $n=0,1,2,\dots$; see Refs.\ \cite{Luk07,Per07,Nat14}. 
The above, together with Eq.\ (\ref{eq:eps_i_ky}), implies the wavenumber
quantization 
\begin{equation}
k_{y}^{(n)}=\pm\frac{\sqrt{2\left(1-\gamma^{2}\right)^{3/2}n}}{\gamma{}\,l_{B}}.
\end{equation}
We further notice that the zero mode ($n=0$) lacks the additional twofold
degeneracy of higher modes ($n>0$).

\begin{figure}[!t]
\centering\includegraphics[width=8cm]{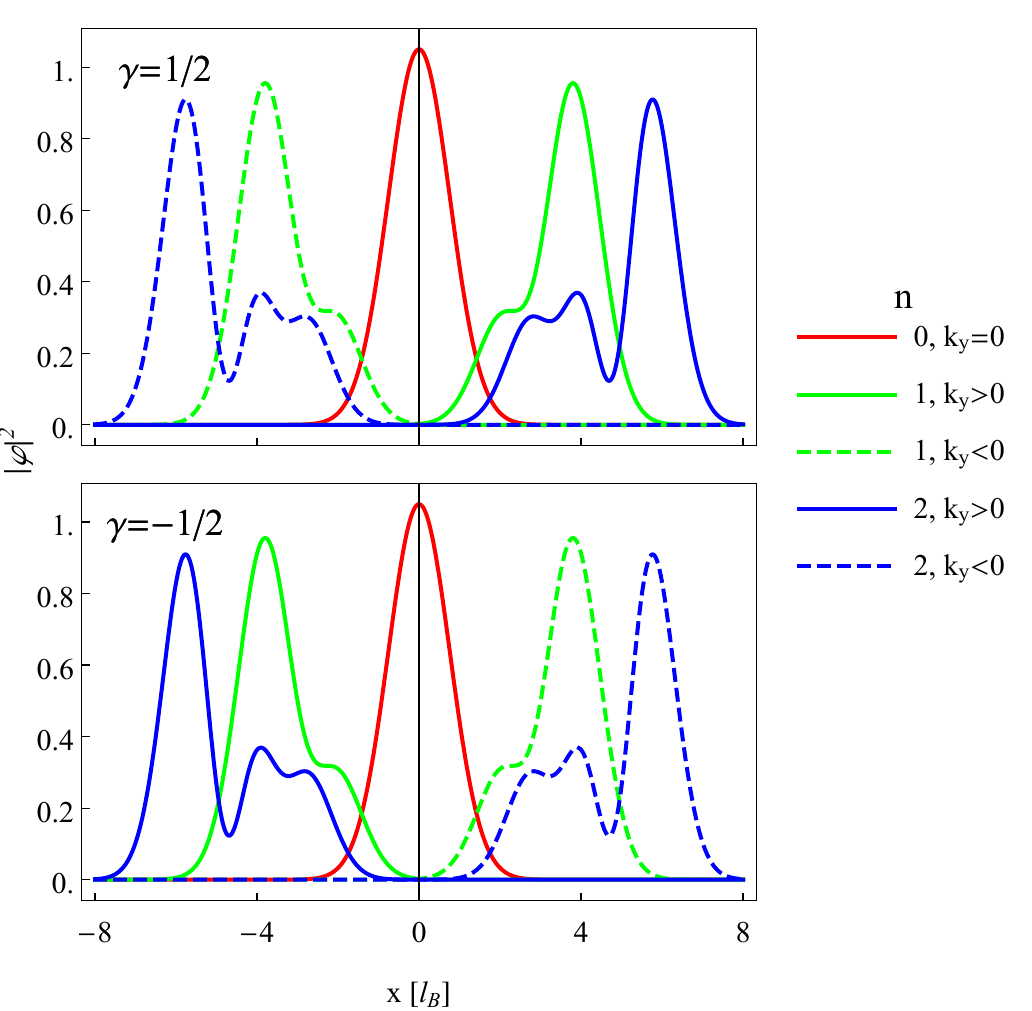}

\caption{ \label{fig:Probability-density}
  Probability density $\left|\varphi\left(x\right)\right|^{2}$ for
  $\xi=1$ (the $K$ valley), $B=1\,$T, 
  and $\gamma=1/2$ (top) or $\gamma=-1/2$ (bottom). 
}
\end{figure}

\begin{figure}[!t]
\centering\includegraphics[width=8cm]{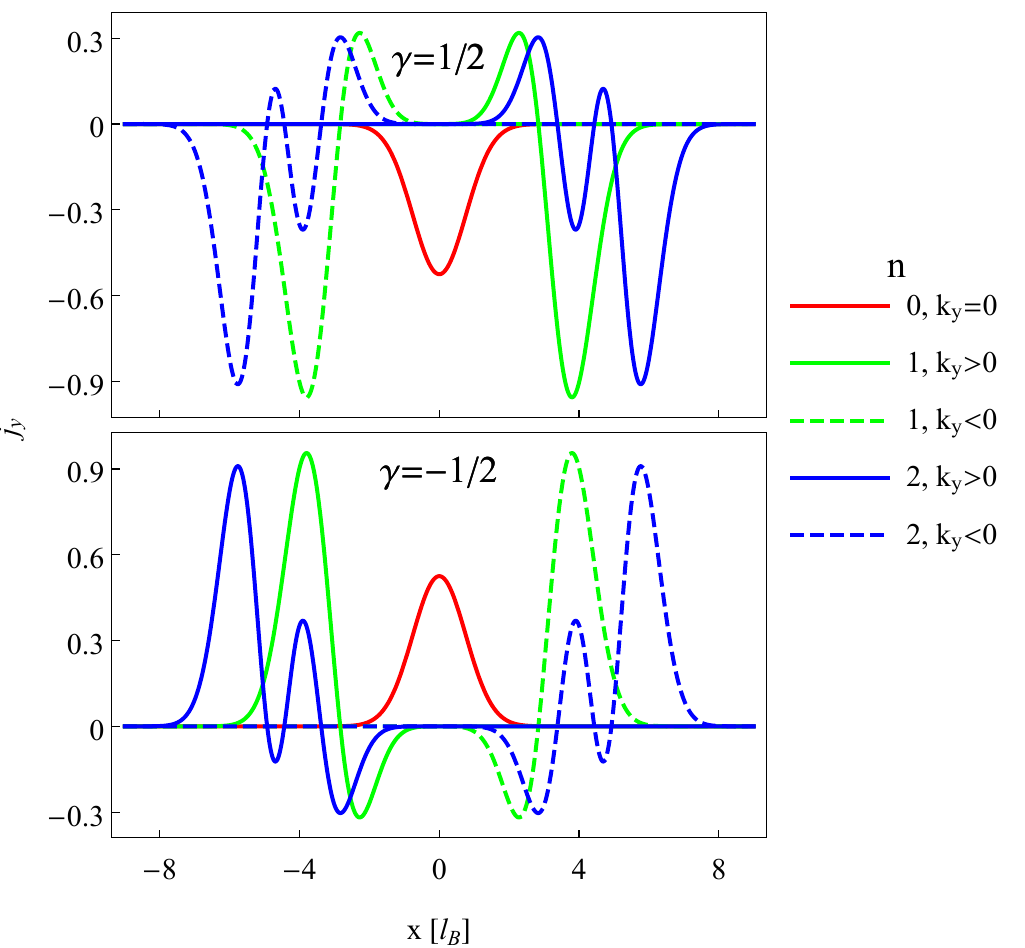}

\caption{ \label{fig:Current-density}
  Current density $j_{y}(x)$ for $\xi=1$ (the $K$ valley), $B=1\,$T,
  and $\gamma=1/2$ (top) or $\gamma=-1/2$ (bottom).
}
\end{figure}

Explicite forms of wavefunctions, given above by Eqs.\ (\ref{eq:varphinon0})
and (\ref{eq:HolyGrail}), 
allows one to calculate the probability density $\left|\varphi\left(x\right)
\right|^{2}$ (see Fig.\ \ref{fig:Probability-density}) 
as well as the local current density
$j_{y}=\varphi^{\dagger}\left(x\right)\cdot\sigma_{y}\cdot\varphi\left(x\right)$
(see Fig.\ \ref{fig:Current-density}).

As we have neglected the mass term throughout this Appendix, the physical
quantities displayed in Figs.\ \ref{fig:Probability-density} and
\ref{fig:Current-density} are same for both valleys, $K$ and $K'$, 
indicated by $\xi=1$ or $\xi=-1$ (respectively) in Eq.\ (\ref{eq:cartH}).
Also, the probability density $\left|\varphi\left(x\right)\right|^{2}$
is affected by the direction of electric field, indicated by 
$\mbox{sgn}\,{\mathcal{E}} \equiv \mbox{sgn}\,\gamma$, only in a~way 
that the two solutions for $n>0$, characterized by
opposite wavenumbers ($k_y$ and $-k_y$) are exchanged upon 
$\gamma\rightarrow{}-\gamma$, see  Fig.\ \ref{fig:Probability-density}. 
In contrast, the current density $j_y(x)$ also changes sign upon
$\gamma\rightarrow{}-\gamma$, see Fig.\ \ref{fig:Current-density}.
Revisiting the derivation for $B<0$, one quickly can find that
$\left|\varphi\left(x\right)\right|^{2}$ and $j_y(x)$ are affected by
the magnetic field inversion ($B\rightarrow{}-B$) at fixed $\gamma$  
in the same way as by the electric field inversion
($\gamma\rightarrow{}-\gamma$) at fixed $B$. 

Another striking feature of the results presented in Fig.\
\ref{fig:Current-density} is that for either the $n=0$ or $n>0$ modes, the
total current (integrated over $x$) flows in one direction only, determined 
by the signs of ${\cal E}$ and $B$.
For $n>0$, this can be attributed to the fact that solutions with $k_y>0$ 
and $k_y<0$ are localized at the opposite sides of a~p-n interface, resulting
in the {\em same} sign of the group velocity.
For $n=0$, the solution given by Eq.\ (\ref{eq:HolyGrail}) can be regarded
as a~linear combination of edge states from both sides of the interface,
for which the current density is centered precisely at the interface line
(as depicted scematically in Fig.\ \ref{qhallfig} in the main text).

\begin{figure}[!t]
\centering\includegraphics[width=\linewidth]{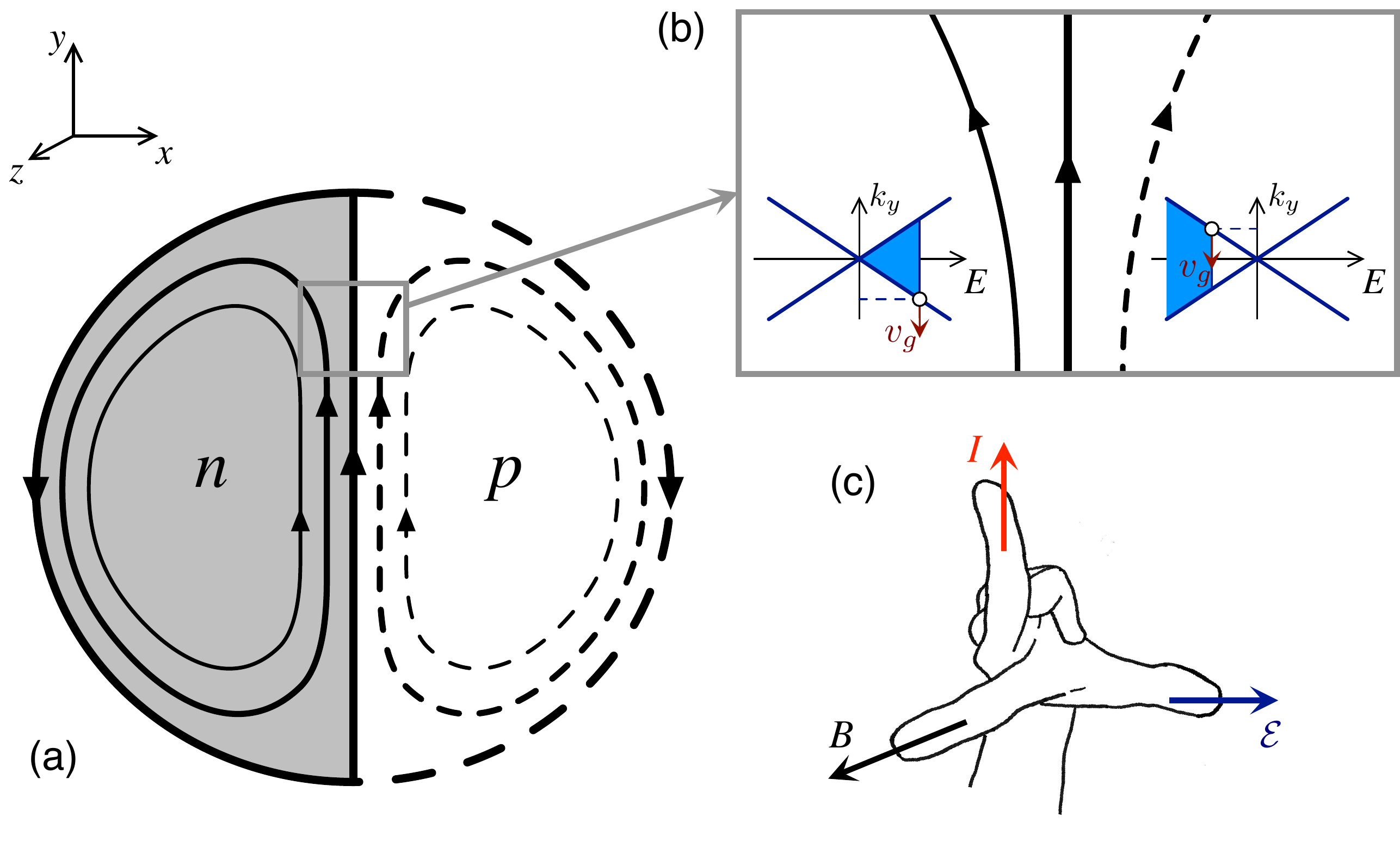}

\caption{ \label{fig:Lhand}
  (a) Edge states in a~finite disk-shaped graphene sample with 
  a~p-n interface in the quantum Hall regime (schematic).
  (b) A~zoom-in of the interface region with band structure schemes.
  Open circles mark the two states on the Fermi level with opposite wavenumbers
  ($k_y<0$ and $k_y>0$) having equal group velocities ($v_g$).
  (Notice that the direction electric current is {\em opposite} to $v_g$ as
  the electron charge is $-e$.)
  (c) A~version of the Fleming's left hand rule showing mutual relation between
  directions of the magnetic field $B$, the in-plane electric field ${\cal E}$,
  and the current $I$ near the interface.
  The coordinate system used in Appendix~\ref{sec:Appendix-B:-Wavefunctions}
  is also shown. 
}
\end{figure}

We comment now the relation between solutions for an infinite plane with
trajectories depicted in Fig.\ \ref{snakefig}.

The snake states (bottom panel in Fig.\ \ref{snakefig}) can be represent as
linear combinations of the solutions with $n=0$ and $n>0$, having a~property
that the full combination propagates in the same direction as each of its
components. 
On the other hand, classical trajectores propagating in the direction
(approximately) perpendicular to the interface (top panel in
Fig.\ \ref{snakefig}) represent finite-size effects having no analogs
in an infinite plane.
Most intriguing are the trajectories depicted in the middle panel of
Fig.\ \ref{snakefig}, propagating in {\em both} directions along
the interface.
Formally, 
this is possible since the total current, considered as a~quadratic form,
is neither positively nor negatively defined, and thus a~generic quantum
state composed of eigenstates with different $n$-s may also carry the
current in opposite direction then each of the components. 

In real sample of a~finite size, edge states associated with a~p-n junction
derived in this Appendix are always accompanied by edge states close to
a~physical system boundary transporting the charge in opposite direction,
see Fig.\ \ref{fig:Lhand}(a).
When a~disk-shaped sample is clamped with circular electrodes, forming
the Corbino setup, edge currents are elliminated by the outer lead and
the schematic current distribution for the lowest modes, visualized
in Fig.\ \ref{fig:Lhand}(b), may be closely reproduced by the physical
current density (see Fig.\ \ref{fig:RhoJ3pan}(a) in the main text). 
Remarkably, the familiar Fleming's left hand rule, relating the directions
of the current, the magnetic field, and the charge displacement
(or the in-plane electric field) has also a~version for graphene 
p-n junction in the quantum Hall regime, see Fig.\ \ref{fig:Lhand}(c).

\section{Mass confinement and the valley separation 
  \label{sec:Appendix-C:-Mass}}

We argue here that the mechanism behind spatial separation of currents
in different valleys, appearing for a~nonzero mass term (see Fig.\
\ref{fig:RhoJ3pan}(b) in the main text), can essentially be understood by
analyzing
the zero-energy wavefunction in the presence of infinite mass confimenent
proposed in the seminal work by Berry and Mondragon \cite{Ber87}.

In the absence of electric field (${\cal E}=0$), a~general zero-energy
solution of Eq.\ (\ref{DiracEqB}) for $\xi=1$ (the $K$ valley)
can be written as \cite{Pra07}
\begin{equation}
\label{pradazero}
  \Phi_{0,k_y,\xi=1}^{[{\cal E}=0]}(x) =
  C_1\left[\begin{array}{c}
     0  \\  e^{-\chi^2/2}
  \end{array}\right]
  + C_2\left[\begin{array}{c}
     e^{\chi^2/2}  \\  0 
  \end{array}\right],
\end{equation}
with $C_1$ and $C_2$ being arbitrary complex numbers, and
$\chi=l_{B}^{-1}x+l_{B}k_{y}$ again.
For $\xi=-1$ (the $K'$ valley), the two basis solutions on the right-hand
side of Eq.\ (\ref{pradazero}) have interchanged spinor components.

Neglecting the intervalley scattering, one can show that confinement of the
carriers in a~bounded domain implies zero outward current at any point
of the boundary at each valley ($\xi=\pm{}1$), namely
\begin{equation}
\label{zerocurrent}
j_{{\bf n}(\alpha)}=
  \left\langle{}\xi\sigma_x\cos\alpha+\sigma_y\sin\alpha
  \right\rangle_{\Phi_\xi}=0,
\end{equation}
where ${\bf n}(\alpha)=(\cos\alpha,\sin\alpha)$ is the unit vector normal
to the boundary, and the spinor wavefunction $\Phi_\xi=\left(
\Phi_{\xi,A},\Phi_{\xi,B}\right)^T$. 
Eq.\ (\ref{zerocurrent}) can be rewritten as
\begin{equation}
\xi\cos\alpha{}\,\mbox{Re}\left(\Phi_{\xi,A}^\star\Phi_{\xi,B}\right) +
  \sin\alpha\,\mbox{Im}\left(\Phi_{\xi,A}^\star\Phi_{\xi,B}\right)=0,
\end{equation}
which is equivalent to
\begin{equation}
 \label{xiphiaphib}
 \left(\frac{\Phi_{\xi,B}}{\Phi_{\xi,A}}\right)^\xi = i{\cal B}\exp(i\alpha),
\end{equation}
where ${\cal B}$ is real and depends on the physical nature of the confinement
\cite{Ber87}. 

Infinite mass confinement at $x=0$, restricting the wavefunction to the right
hemiplane ($x>0$), corresponds to ${\cal B}=1$ and $\alpha=\pi$ in Eq.\
(\ref{xiphiaphib}) and leads to the boundary condition
\begin{equation}
\left.\Phi_{\xi,A}\right|_{x=0} = i\xi\left.\Phi_{\xi,B}\right|_{x=0}. 
\end{equation}
Subsequently, the coefficients in Eq.\ (\ref{pradazero}) follow
\begin{equation}
\label{infma1c2}
  C_{\xi,2}=i\xi{}C_{\xi,1}\exp\left(-k_y^2l_B^2\right). 
\end{equation}
The vertical current density for the zero-energy solution is
\begin{equation}
  j_y =
  \left\langle{}\sigma_y\right\rangle_{\Phi_{0,k_y,\xi}^{[{\cal E}=0]}}=
  -2\xi{}|C_{\xi,1}|^2\exp\left(-k_y^2l_B^2\right), \label{averjy0}
\end{equation}
where
the last equality follows from Eq.\ (\ref{infma1c2}). 

Clearly, the uniform current in Eq.\ (\ref{averjy0}) changes sign upon
the valley exchange ($\xi\rightarrow{}-\xi$), providing a~qualitative
understanding of the valley separation, as the effect associated with
zero-energy mode should overrule the effects originating from higher
modes for a~generic system close to the charge-neutrality point
(allowing for $|k_y{}l_B|\ll{}1\,$).



\begin{thebibliography}{}
\bibitem{claquafoo}
For $r_c\lesssim{}l_B$, quantum effects dominate and may fully elliminate
the moderate-field range (see midle panel in Fig.~\ref{snakefig}) for smaller
disks.
As $r_c$ is related to the carrier concentration, 
particular scenario of the~classical-to-quantum transition also
depends on the electrostatic potential profile.
\bibitem{zeemafoo}
For $B=1\,$T the Zeeman splitting is $\Delta{}E_Z = g\mu_B{}B
\approx{}0.1\,$meV [with $g\approx{}2$ and $\mu_B=e\hbar{}/(2m_e)$ the Bohr
magneton] and cannot affect the filter operation appearing at the potential
and mass-term amplitudes of $V\sim{}M\gtrsim{}10\,$meV.
\end{thebibliography}
\end{document}